\pdfoutput=1
\documentclass[journal = jpccck, manuscript = article]{achemso}
\setkeys{acs}{usetitle = true}
\usepackage{graphicx}
\usepackage{float}
\usepackage{achemso}
\usepackage{natbib}
\usepackage{multirow}
\usepackage{wrapfig}
\usepackage{times}
\usepackage{booktabs}
\usepackage{rotating}
\usepackage{amsmath}
\usepackage{mathptmx}
\usepackage[version=3]{mhchem}  
\usepackage{url}

\mathchardef\mhyphen="2D

\title{CO-Induced Restructuring on Stepped Pt Surfaces: A Molecular
  Dynamics Study}

\author{Joseph R. Michalka}
\affiliation[University of Notre Dame]{Department of Chemistry and
  Biochemistry\\ University of Notre Dame\\251 Nieuwland Science
  Hall\\Notre Dame, Indiana 46556}
\alsoaffiliation[Southwest Baptist University]{Department of Chemistry and Physics\\Southwest Baptist University\\Wheeler Science Facility\\Bolivar, Missouri 65613}
\author{Andrew P. Latham}
\author{J. Daniel Gezelter}
\email{gezelter@nd.edu}
\phone{(574) 631-7595}
\affiliation[University of Notre Dame]{Department of Chemistry and
  Biochemistry\\ University of Notre Dame\\251 Nieuwland Science
  Hall\\Notre Dame, Indiana 46556}

\begin{document}

\begin{abstract}
  The effects of plateau width and step edge kinking on carbon
  monoxide (CO)-induced restructuring of platinum surfaces were
  explored using molecular dynamics (MD) simulations. Platinum
  crystals displaying four different vicinal surfaces [(321), (765),
  (112), and (557)] were constructed and exposed to partial coverages
  of carbon monoxide. Platinum-\ce{CO} interactions were fit to recent
  experimental data and density functional theory (DFT) calculations,
  providing a classical interaction model that captures the atop
  binding preference on \ce{Pt}. The differences in \ce{Pt\bond{-}Pt}
  binding strength between edge atoms on the various facets were found
  to play a significant role in step edge wandering and reconstruction
  events. Because the mechanism for step doubling relies on a
  stochastic meeting of two wandering edges, the widths of the
  plateaus on the original surfaces was also found to play a role in
  these reconstructions. On the Pt(321) surfaces, the CO adsorbate was
  found to assist in reordering the kinked step edges into straight
  \{100\} edge segments.
\end{abstract}
\newpage

\section{Introduction}
Industrial catalysts often consist of supported metal nanoparticles or
high-index metal surfaces. Both of these materials have a high density
of undercoordinated atoms which have been shown to be active for
catalytic
reactions.\cite{Jeong:1999aa,Larsen:1999aa,Shi:2016kl,Tao:2016cr}
Catalytic activity on low-energy metal surfaces, primarily the (111),
(110), and (100) facets, is only indirectly applicable to industrial
conditions.\cite{Ertl:1977cg,Levine:1986ly,Wong:1991ys,Greeley:2002ud}
With new experimental techniques like high pressure scanning tunneling
microscopy (HP-STM) and ambient pressure X-ray photoelectron
spectroscopy (AP-XPS), high index surfaces can now be explored at
higher temperatures and pressures.  This allows for a more complete
understanding of the surface structures of industrial
catalysts.\cite{Stamenkovic:2007kk,Stephens:2011bv,Mohsenzadeh:2015kx}
Roughened surfaces and nanoparticles are of particular interest
because morphology and stability at high temperatures and pressures
depends on the presence and identity of
adsorbates.\cite{Tao:2008aa,Tao:2010aa,Tao:2011aa,Michalka:2013aa,Michalka:2015aa,Kim:2016cr,Eren:2016qt}
Under reaction conditions, many of these catalysts have also been
observed to reconstruct, drastically changing the surface of the
catalyst and altering the activity and selectivity of the
material.\cite{Balmes:2016uf,Kim:2016cr,Eren:2016qt}

Carbon monoxide (\ce{CO}) oxidation on platinum (\ce{Pt}) surfaces has
been studied extensively as a model catalyst. One system that has
received significant attention is the \ce{Pt}(557)
surface.\cite{Tao:2010aa,Tao:2011aa,Michalka:2013aa,Kim:2016cr} Tao
{\em et al.}, using HP-STM, AP-XPS, and density functional theory
(DFT) calculations, observed that by introducing \ce{CO} to the
system, the stepped \ce{Pt} surface would undergo a step-doubling
reconstruction yielding steps that were twice as high and plateaus
that were twice as wide as the original surface.\cite{Tao:2010aa} If
the system was maintained at these conditions, the appearance of
triangular nanoclusters bridging the steps were also observed. After
removing \ce{CO} from the system, it was found that these changes were
reversible, as the \ce{Pt}(557) surface was recovered. It was
suggested that this reconstruction was caused by the strong repulsive
interactions between adsorbed \ce{CO} molecules. This system has also
been explored with molecular dynamics simulations where a possible
mechanism, involving the large \ce{CO}-\ce{CO} quadrupolar repulsion
from adjacent atop adsorption sites near step edges, was proposed
to explain the step-doubling process.\cite{Michalka:2013aa}

The strong interaction of \ce{CO} with \ce{Pt} has been explored on
different systems and similar results have been observed. Ferrer {\em
  et al.}, examining \ce{Pt}(997), observed both \ce{O} and \ce{CO}
induced step-doubling on the surface. They also saw that the presence
of step-doubling was correlated with a factor of three increase for
\ce{CO} oxidation rates.\cite{Balmes:2016uf} Park {\em et al.}, again
exploring the \ce{Pt}(557) system exposed to \ce{CO}, saw that the
ordering of the triangular nanoclusters is very sensitive to
temperature and can undergo its own reversible restructuring separate
from the doubling already observed on this surface.\cite{Kim:2016cr}

Similarly, Eren {\em et al.} found that a flat \ce{Cu}(111) surface,
when exposed to \ce{CO}, also underwent significant reconstruction,
forming small nanoclusters on the surface that were highly active for
the oxygen reduction reaction (ORR).\cite{Eren:2016qt} In all of these
cases, the presence of adsorbates led to large-scale reconstructions
(often reversible) of the surface. However, the dependence of the
reconstruction dynamics on the originally displayed surface structure
is still unclear.

This paper attempts to answer the question of how initial surface
morphology affects reconstruction pathways by examining the effect of
\ce{CO} on the reconstruction of various \ce{Pt} facets, specifically
focusing on the effect of edge type and plateau width. The
\ce{Pt}(321)~\cite{Mcclellan:1981aa,Bray:2014aa,Bray:2011aa},
\ce{Pt}(112)~\cite{Xu:1995aa,Yates:1995aa}, \ce{Pt}(765), and
\ce{Pt}(557)~\cite{Michalka:2015aa,Michalka:2013aa,Tao:2010aa,Kim:1997aa}
vicinal surfaces were chosen as they represent a mixture of straight
and kinked atomic steps as well as narrow and wide plateaus. Based on
the proposed mechanism in Ref. \citenum{Michalka:2013aa}, it is
hypothesized that the width of the plateau and the roughness of the
step edge will both contribute to the reconstruction on these
surfaces. These systems were modeled using classical force fields that
balance chemical accuracy with computational efficiency. At room
temperature, the structural changes occur on timescales of 10-100s, so
molecular dynamics simulations with elevated temperatures are
necessary to observe the evolution of these systems.

\section{Methodology}
Modeling surface reconstructions of solid/gas interfaces requires
relatively large systems and long timescales. To observe events such
as step doubling and step wandering, systems need to contain on the
order of 10\textsuperscript{3}--10\textsuperscript{4} atoms that are
observed for tens to hundreds of nanoseconds. The large number of
electrons associated with metallic interfaces makes modeling these
interfaces with {\em ab initio} molecular dynamics
(AIMD),\cite{Kresse:1993aa,Kresse:1993ab,Kresse:1994aa} Car-Parrinello
methods,\cite{Car:1985aa,Izvekov:2000aa,Guidelli:2000aa} or quantum
mechanical potential energy surfaces intractable.  However, the
interactions between metal atoms are poorly represented by pairwise
interactions.  Therefore, the Embedded Atom Method (EAM)
\cite{Foiles:1986ky} was used to model the \ce{Pt\bond{-}Pt}
interactions as it can effectively handle large numbers of atoms while
including non-pairwise effects. For modeling \ce{CO}, a three site
model developed by Straub and Karplus was
utilized.\cite{Straub:1991aa} This model treats the \ce{CO} as a rigid
body and was chosen primarily for its accurate description of the
large linear quadrupole of \ce{CO}. The \ce{Pt\bond{-}CO} interaction
was modeled to match a combination of recent experimental
\cite{Yeo:1997th,Ertl:1977cg,Kelemen:1979ad} and theoretical
\cite{Deshlahra:2012aa,Korzeniewski:1986aa,Beurden:2002aa,Deshlahra:2009aa,Feibelman:2001aa,Mason:2004aa}
data.

\subsection{Platinum-Platinum Interactions}
Since force fields with only pairwise interactions do not treat
transition metal cohesive energies correctly, several methods have
been developed that describe metallic interactions utilizing
non-pairwise additive functions of the local electron density. Some of
these methods include the embedded atom (EAM)
\cite{Foiles:1986ky,Daw:1984aa,Johnson:1989aa,Daw:1989aa,Plimpton:1992aa,Lu:1997aa,Alemany:1998aa,Voter:1994aa},
the Finnis-Sinclair~\cite{Finnis:1984aa,Sutton:1990aa} and the
quantum-corrected Sutton-Chen methods~\cite{Qi:1999aa}. In general,
these methods treat a metal atom as a positive core surrounded by a
radially decaying charge density representing the valence electrons.
Computing the energy for placing atom $i$ in a specific location
requires computation of the electron density contributed by all other
metallic atoms in the system,
\begin{equation}
\rho_{i} = \sum_{j \ne i}\rho_{j}(r_{ij}).
\label{eq:density}
\end{equation}
$\rho_{j}(r_{ij})$ is the valence electron density contributed by atom
$j$ at site $i$.  The potential energy of atom $i$ is provided by a
functional, $F_i \left[\rho_{i}\right]$, which describes the
attractive energy between the positive core of the atom and the
background electron density contributed by the surrounding metal,
\begin{equation}
V_{i} = F_i \left[\rho_{i}\right] + \sum_{j \ne i} \phi_{ij}(r_{ij}),
\label{eq:eam}
\end{equation}
while $\phi_{ij}(r_{ij})$ represents the pairwise repulsion between
the positively charged cores. Potentials similar to the forms in
Eqs. (\ref{eq:density}) and (\ref{eq:eam}) are used in the EAM,
Finnis-Sinclair, and Quantum Sutton Chen models. These models have
been used for a variety of theoretical calculations of bulk and
nanoparticle properties,
\cite{Mishin:1999aa,Chui:2003aa,Wang:2005aa,Medasani:2007aa}
melting,\cite{Belonoshko:2000aa,Sankaranarayanan:2005aa,Sankaranarayanan:2006aa}
fracture, \cite{Shastry:1996aa,Shastry:1998aa,Mishin:2001aa} crack
propagation, \cite{Becquart:1993aa} and alloying
dynamics.\cite{Shibata:2002aa,Mishin:2002aa,Zope:2003aa,Mishin:2005aa}
As the EAM parameterization includes second and third nearest-neighbor
interactions, it is particularly suited to interfaces that deviate
from low-energy (111) surfaces, and the ``u3'' EAM fits are utilized
in this work.\cite{Foiles:1986ky}

\subsection{Surface Models}
Vicinal \ce{Pt} surfaces were generated by slicing an ideal FCC
lattice along planes corresponding to specific Miller indices
$(hk\ell)$.  Two perpendicular cuts were used to generate an
orthorhombic slab, and the desired plane was oriented along the $z$
axis of the simulation box, which is periodic in all three dimensions.
For each of the simulated surfaces, Pt(321), Pt(112), Pt(765), and
Pt(557), ideal surfaces were doubled in either the $x$ or the $y$
direction.  These systems are identified as LS, for systems with
longer steps, and MS, for systems with more steps. The system
dimensions, number of atoms, and number of surface atoms are
enumerated in Table \ref{tab:dimensions}.  In each case, the box was
extended along the $z$-axis to a length of 100~\AA~to allow for a
sufficient gas phase layer between periodic replicas of the system.
Note that each simulation cell displays two of the $\{hk\ell\}$ facets
to the gas, one in the positive $z$ direction and one facing the
negative $z$ axis.
\begin{sidewaystable}
\caption{Surface Models}
\centering
\begin{tabular}{c c c c c c c c}
\hline\hline
Pt & Atoms & Surface & 
\multicolumn{2}{c}{dimensions (\AA)} 
& Slab thickness & Surface Energy & Plateau Width \\
Surface & & Atoms & MS & LS & (\AA) & (J/$\textrm{m}^2$) & (\AA) \\
\hline
(321) & 4440 & 720 & $71.7 \times 47.7$ & $35.8 \times 95.5$  & 18.85 &1.76 & 5.90 \\
(112) & 4608 & 768 & $32.9 \times 108.5$  & $65.8 \times 54.3$ & 18.40 &1.67 & 6.79 \\
(557) & 3888 & 720 & $110.6 \times 24.8$ & $55.3 \times 49.5$ & 18.52 &1.55 & 13.79 \\
(765) & 3744 & 792 & $28.7 \times 100.5$  & $58.4 \times 50.2$  & 19.24& 1.53 & 16.79\\
\hline
\hline
\end{tabular}
\label{tab:dimensions}
\end{sidewaystable}

Surface energies were calculated for every system that was studied, as
well as for the low-index \ce{Pt}(111), \ce{Pt}(100), and \ce{Pt}(110)
surfaces. To perform these calculations, systems were created such
that they were periodic in all directions $(x,y,z)$. The desired
surface was then exposed by extending the box in the $z$-direction a
sufficient distance to prevent long-range interactions between the
exposed facets. The change in the potential energy that resulted was
used to calculate the surface energy. The surface energies calculated
for the low-index surfaces match previous values calculated using the
same potential.\cite{Foiles:1986ky} Note that for the low-index
facets, these values are a factor of $\sim 1.3$ larger than surface
energies obtained via
experiments~\cite{Tyson:1977xe,De-Boer:1988tg,Galeev:1980pt} and
DFT~\cite{Vitos:1998qq}, but the ratios of surface energies are
preserved relative to experimental ratios. The EAM surface energies
for the higher index facets are also shown in Table
\ref{tab:dimensions}, and by analogy with the low index facets, these
are expected to be a factor of 1.3 larger than experimental surface
energies for the same facets. We note that the surface energies of
these systems are inversely correlated with the plateau widths. Larger
plateaus correspond to larger regions of exposed low energy (111)
facets.


\subsection{Carbon Monoxide}
The large linear quadrupole moment of \ce{CO} is believed to play a
role in the reconstruction of
\ce{Pt}(557).\cite{Tao:2010aa,Michalka:2013aa} The model of Karplus
and Straub effectively reproduces that quadrupole.\cite{Straub:1991aa}
This model consists of a rigid body comprising three sites. Two sites
describe the \ce{C} and \ce{O}, both with a partial negative charge,
while the third site (\ce{M}) is massless and carries a large
neutralizing positive charge.  The \ce{M} site is located at the
center of mass of the molecule.  The parameters for this model are
shown in Table \ref{tab:parameters}.  These parameters produce a
molecule with a small dipole moment (0.35 D) while keeping the linear
quadrupole (-2.40 D\AA) close to both experimental calculations (-2.63
D\AA)\cite{Chetty:2011aa} and quantum mechanical predictions (-2.46
D\AA).\cite{Noro:2000aa}

\begin{table}[ht]
\caption{Positions, Lennard-Jones parameters ($\sigma$ and $\epsilon$) and charges for \ce{CO\bond{-}CO} interactions}
\centering
\begin{tabular}{c c c c c}
\hline\hline
 & z (\AA) & $\sigma$ (\AA) & $\epsilon$ (kcal/mol) & q ($e^{-}$) \\
 \hline
 C & -0.6457 & 3.83 & 0.0262 & -0.75 \\
 O & 0.4843 & 3.12 & 0.1591 & -0.85 \\
 M & 0.0 & & & 1.6 \\
 \hline\hline
 \end{tabular}
\label{tab:parameters}
\end{table}

\subsection{Platinum-Carbon Monoxide Interactions}
Extensive experimental
\cite{Ertl:1977cg,Kelemen:1979ad,Yeo:1997th,Hort:2015jk,Myshlyavtsev:2015bv,Kim:2016cr}
and theoretical
\cite{Korzeniewski:1986aa,Feibelman:2001aa,Beurden:2002aa,Mason:2004aa,Deshlahra:2009aa,Deshlahra:2012aa}
work on \ce{Pt\bond{-}CO} systems allows for significant data for
parameterizing a force field for the \ce{Pt}-\ce{CO} interaction.  The
\ce{Pt\bond{-}CO} model used here has evolved from our previous
models\cite{Michalka:2013aa,Michalka:2015aa} in order to improve the
difference in energy between the preferred atop and bridge sites
described by recent DFT calculations.\cite{Deshlahra:2012aa} The
current model uses a repulsive Morse potential to represent the
\ce{Pt\bond{-}O} interaction, while previous implementations have used
a standard ``shifted'' Morse
potentials.\cite{Korzeniewski:1986aa,Michalka:2013aa} The repulsive
Morse potential prevents unphysical \ce{O}-first binding to the
surface.  Combining the Lennard-Jones \ce{Pt\bond{-}C} interaction
with the repulsive Morse model for the \ce{Pt\bond{-}O} interaction,
the resulting potential,
\begin{equation}
V_\mathrm{PtCO} = 4\epsilon \left(
 \left( \frac{\sigma}{r_\mathrm{PtC}} \right)^{12} -
 \left( \frac{\sigma}{r_\mathrm{PtC} }\right)^{6} \right) + D_{e} e^{-2\gamma(r_\mathrm{PtO}-r_{e})}.
\end{equation}

Table \ref{table:pt-co} shows the parameters for the \ce{Pt}-\ce{CO}
interaction, while Table \ref{table:sites} provides the binding
energies for the various (111) binding sites based on these parameters
and compares them to DFT\cite{Deshlahra:2012aa} and experimental
data.\cite{Ertl:1977cg,Kelemen:1979ad}
\begin{table}[ht]
\caption{Parameters for Pt-CO Interaction}
\centering
\begin{tabular}{lll}
\hline\hline
\multicolumn{3}{l}{Pt-C} \\
 & $\sigma$   & 1.68 \AA \\
 & $\epsilon$ &  52.25 kcal / mol \\
\hline
\multicolumn{3}{l}{Pt-O} \\
 & $r_e$ & 5.15 \AA \\
 & $D_e$ & 0.03 kcal/mol \\
 & $\gamma$ & 1.7 \AA$^{-1}$ \\
\hline
\hline
\end{tabular}
\label{table:pt-co}
\end{table}

\begin{table}[ht]
\caption{Pt-CO Binding Site Preferences (eV)}
\centering
\begin{tabular}{c c c c}
\hline\hline
& Atop & Bridge & Hollow \\
\hline
This work & -1.49 & -1.36 & -1.32 \\
DFT \cite{Deshlahra:2012aa} & -1.48 & -1.47 & -1.45 \\
Experimental\cite{Ertl:1977cg, Kelemen:1979ad} & -1.43 & & \\
\hline\hline
\end{tabular}
\label{table:sites}
\end{table}

\subsection{Simulation Protocol}

The bare interfaces were run in the canonical (NVT) ensemble, where
the temperature was gradually raised to either 700~K for Pt(321) or
1000~K for Pt(112), Pt(765), and Pt(557). Systems with wider plateaus
are significantly more stable and were run at 1000~K to bring
reconstruction dynamics into the domain of typical simulation times.
In the STM papers (Refs. \citenum{Tao:2010aa} and
\citenum{Kim:2016cr}), CO was present at pressures between
$5 \times 10^{-9}$ and $1.05$ Torr, yielding coverages between 0.5 and
1 ML.  In our simulations, gas phase \ce{CO} molecules, corresponding
to 0.25 ML or 0.5 ML of surface coverage were placed in the vacuum
region and then allowed to adsorb on the surface. The dosed systems
were re-equilibrated for one nanosecond before data collection.  Data
was collected by running the systems in the microcanonical (NVE)
ensemble for 100 ns. Simulations were performed using the open source
molecular dynamics package, OpenMD.\cite{Meineke:2005pt,openmd}

\subsection{Generalized coordination number}
For many catalytic reactions, only a subset of the atoms on a
roughened surface or nanoparticle are catalytically active. While the
coordination numbers of individual atoms can describe their binding
strengths, Calle-Vallejo {\it et al.}  observed that including the
first and second nearest neighbor counts allowed for a more complete
description of an atom's local environment and its catalytic
activity.\cite{Calle-Vallejo:2015qq} They introduced the
\textit{generalized coordination number} (GCN), to describe this
quantity,
\begin{equation}
  \overline{CN}(i) = \sum_{j=1}^{n_i}\frac{cn(j)}{cn_{\textrm{max}}}
  \label{eq:gcn}
\end{equation}
The GCN is an extension of nearest-neighbor analysis where the GCN of
atom $i$, $\overline{CN}(i)$, is calculated from the average of the
coordination numbers, $cn(j)$, of atom $i$'s nearest neighbors ($j$).
The sum is carried out over atom $i$'s first solvation shell, and
$cn_\textrm{max}$ is a normalization term.  For an FCC crystal,
$cn_\textrm{max} = 12$, the bulk coordination number.
\begin{figure}[p!]
  \includegraphics[width=\linewidth]{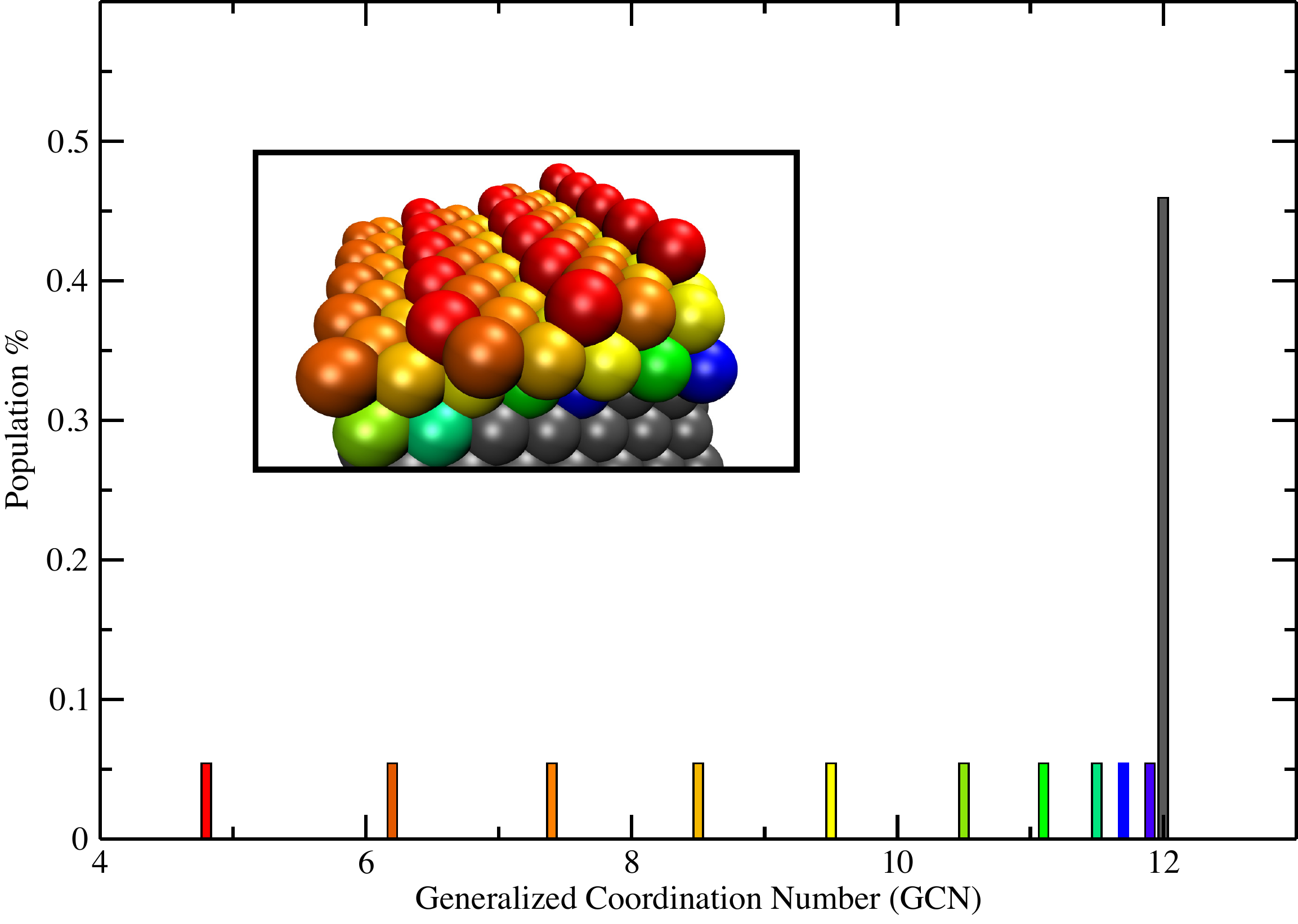}
  \caption{The generalized coordination number of an ideal Pt (321)
    system colored to match the inset structure. Other than the bulk
    (GCN = 12), the ideal (321) surface displays a wide variety of
    coordination environments at its surface. The under-coordinated
    edge atoms (red) have the lowest GCN, where the plateau atoms have
    GCN values around the ideal (111) surface number of 7.5.  Subsurface
    atoms (yellow, greens, and blues) have a larger GCN than the
    surface, but are still less coordinated than the bulk.}
\label{fig:ideal321GCN}
\end{figure}
The concept is further illustrated in Figure \ref{fig:ideal321GCN}
where we see an ideal \ce{Pt}(321) surface color coded to match the
computed GCN values.  The bulk atoms (GCN=12) make up the majority of
the system.  However, there is a wide range of potentially
catalytically active sites on the surface.  Calle-Vallejo {\em et al.}
argued that for the Oxygen Reduction Reaction (ORR) a \ce{Pt} atom
with a GCN of $\sim$8.3 would be the most catalytically active
site. Since an ideal (111) surface in composed of atoms with a GCN of
$7.5=(6\times9 + 3\times12)/12$, this implies that surfaces that
display some amount of concavity may be necessary to achieve the
highest catalytic activities. The GCN distribution for an ideal
\ce{Pt} (557) system is included in the Supporting Information as
Fig. S1 to highlight the differences between kinked
(321) edges and the flat edges of the (557) surface.

\subsection{Step-edge detection}
\begin{figure}[p!]
  \includegraphics[width=4in]{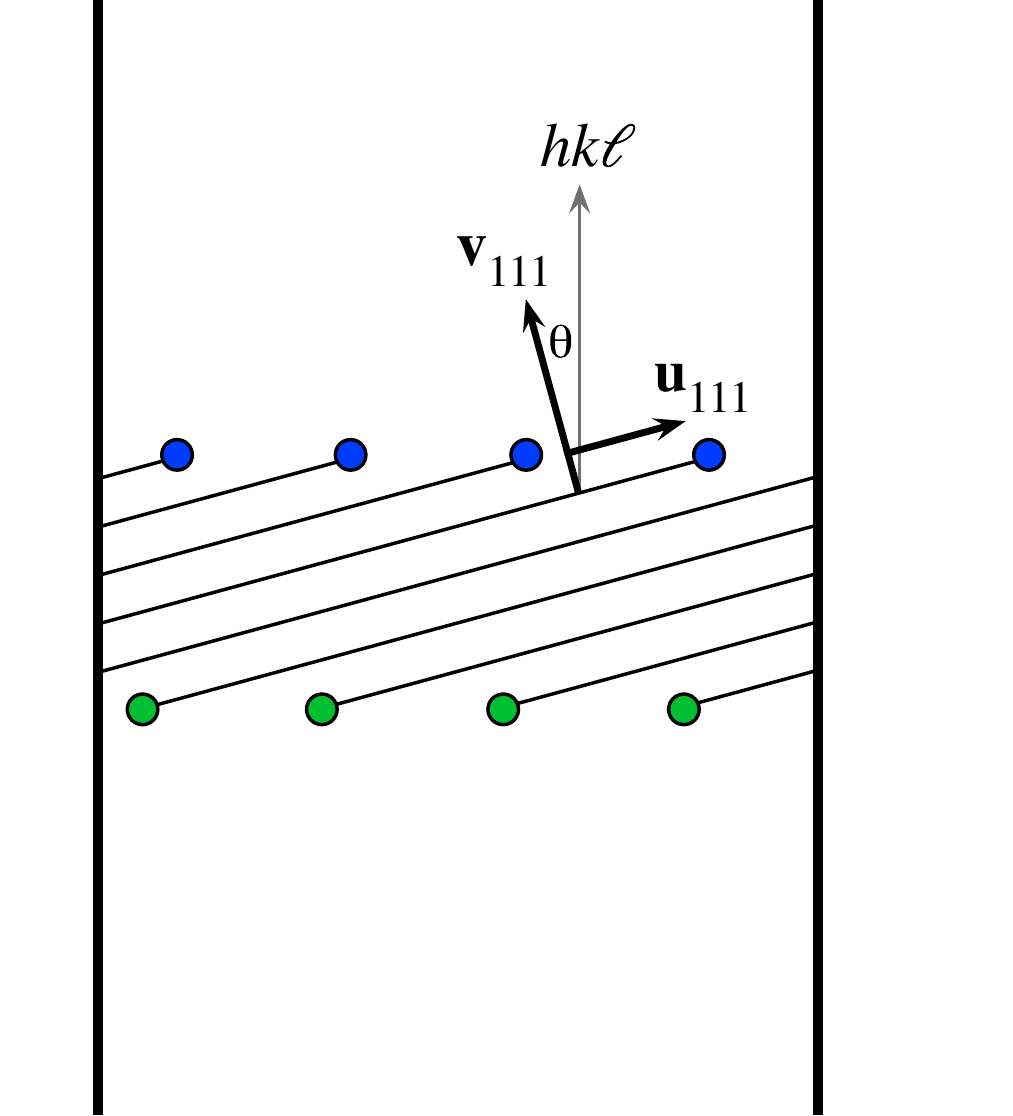}
  \caption{Projection of the $hk\ell$ normal vector onto the $(111)$
    normal vector ($\mathbf{v}_{111}$) to find an angle ($\theta$) for
    the tilt of the step edges relative to the box geometry. Step edge
    atoms are defined by the maximum and minimum $\mathbf{u}_{111}$
    values on each plateau plane. The step edges project into the
    plane of the figure.}
\label{fig:EdgeSketch}
\end{figure}

Because individual steps on the metal surface display the low energy
(111) facet, our method for identifying atoms in step edges divides
the bulk metallic system into a series of (111) planes.  We make use
of the Miller indices ($hk\ell$) of the initial interface, {\em e.g.}
(557) which have been oriented to sit perpendicularly to the $z$-axis
of the simulation cell. The vector normal to the (111) plateaus
defines an angle that the plateaus make with the simulation cell (see
fig. \ref{fig:EdgeSketch}),
\begin{equation}
  \textrm{cos}(\theta) = \frac{h + k + \ell}{\sqrt{3 (h^2 + k^2 + \ell^2)}}
\label{eq:vecDet}
\end{equation}
The surfaces were constructed with the initial step edge lying along
either the $x$- or $y$-axes, so one of the two vector components is
zero. If the system has the initial step edge oriented along the
$y$-axis, the vector normal to the (111) plateaus is
\begin{equation}
\mathbf{v}_{111} = \left( \begin{array}{c} 
\sin \theta \\
0 \\
\cos \theta \end{array} \right).
\label{eq:newNormal}
\end{equation}
Atoms are assigned to a plateau by projecting their instantaneous
coordinates onto the (111) normal vector,
$v_i = \mathbf{r}_i \cdot \mathbf{v}_{111}$.  Histograms of these
projections allow separation of the (111) planes, and each atom is
assigned to a plateau using the instantaneous $v_i$ values.

If the initial step edge runs parallel to the $y$-axis, we may define
a new vector that is perpendicular to both the step edge and the
(111)-normal,
\begin{equation}
\mathbf{u}_{111} = \mathbf{v}_{111} \times \mathbf{\hat{y}}
\end{equation}

The locations of atoms within a single (111) plane are transformed into
the new coordinate system with basis vectors
$\left\{\mathbf{u}_{111}, \mathbf{\hat{y}}, \mathbf{v}_{111}\right\}$.
For a given value of the $\mathbf{\hat{y}}$ coordinate, it is
relatively straightforward to find the atoms with the minimum and
maximum values along the $\mathbf{u}_{111}$ coordinate. These atoms
are shown with blue and green dots in fig. \ref{fig:EdgeSketch}.

In periodic boundary conditions, each of the (111) plateaus exhibits two
edges in the simulation cell, and the top and bottom edge atoms are
collected separately and sorted by their $\hat{y}$ values.  This list
of atoms is a single ``edge'' that can be followed dynamically and
analyzed over the course of a trajectory.

\section{Results}
\subsection{Changes in surface coordination}
\begin{figure}[p!]
  \includegraphics[width=\linewidth]{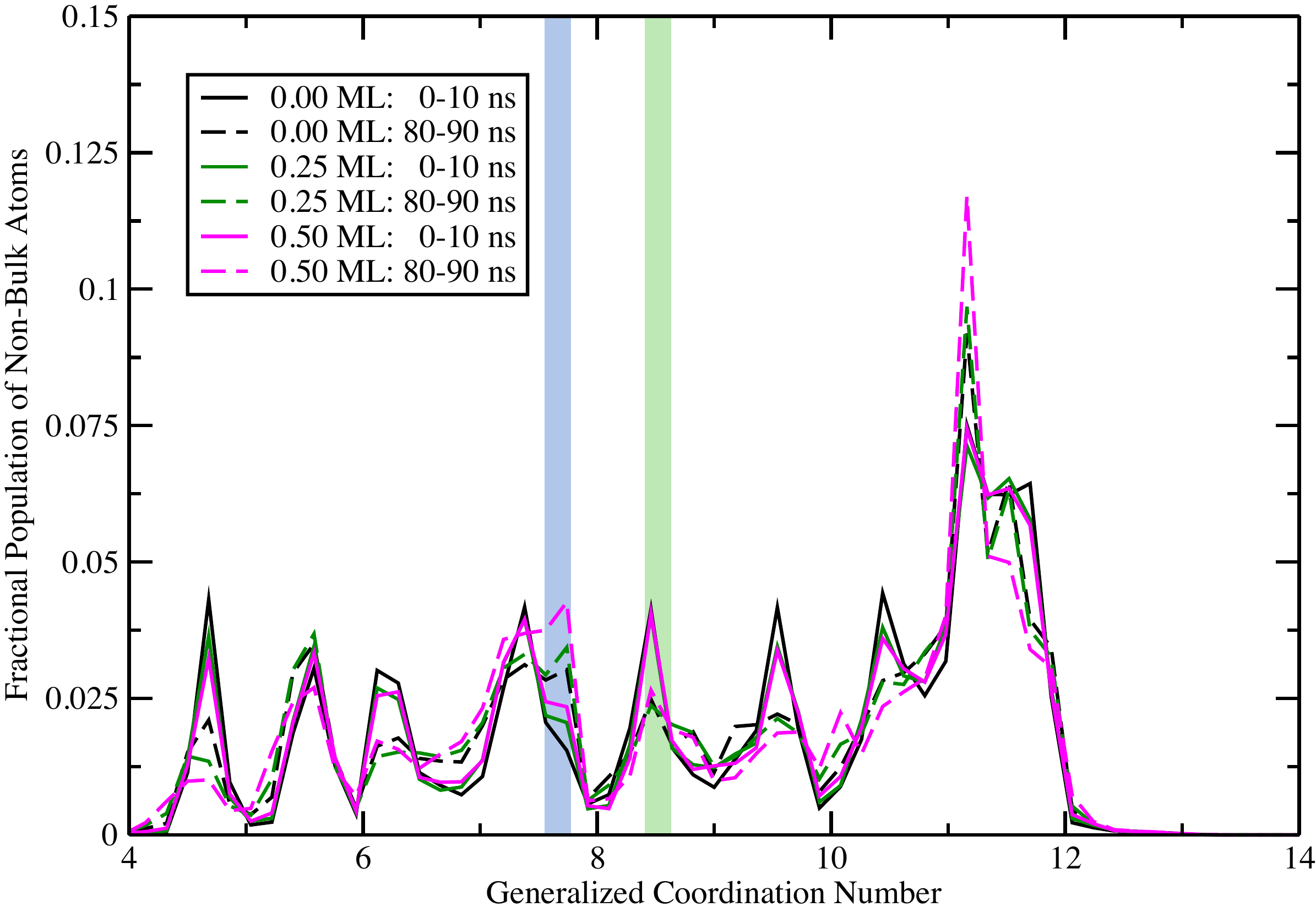}
  \caption{The distribution of generalized coordination numbers (GCN)
    for the near-surface atoms of the \ce{Pt}(321) LS systems.  Solid
    lines represent the averaged GCN during the first 10 ns of
    simulation time, while dashed lines correspond to the last 10 ns.
    The undosed surface (black) exhibits moderate reconstruction,
    while the 0.25 ML (green) and 0.50 ML (magenta) show significantly
    more evolution towards (111) plateaus.  The blue bar is aligned with
    GCN$\sim$7.5 which corresponds to surface atoms in an ideal
    \ce{Pt} (111) surface, while the green bar highlights the region
    identified by Calle-Vallejo {\em et al.} as being especially
    active for ORR activity.\cite{Calle-Vallejo:2015qq}}
\label{fig:LS321GCNF}
\end{figure}

In Figure \ref{fig:LS321GCNF}, we show the generalized coordination
distribution from the \ce{Pt}(321) systems which have been limited to
the surface and subsurface layers, i.e. non-bulk, for three CO
coverages at both the beginning and end of the simulations. The growth
in the peaks around 7.5, highlighted with a blue bar, suggests that
the surface has undergone reconstruction resulting in larger (111)
domains on the surface. The amount of CO present in the system plays a
direct role in this reconstruction. The loss of height in peaks at GCN
$\sim$~4.5 and $\sim$~6.2 suggest that these systems are
reconstructing by displaying more of the lower energy (111) facets.  The
growth in GCN near 11.2 indicates that the subsurface layers are
becoming more bulk-like.

\begin{figure}[p!]
  \includegraphics[width=\linewidth]{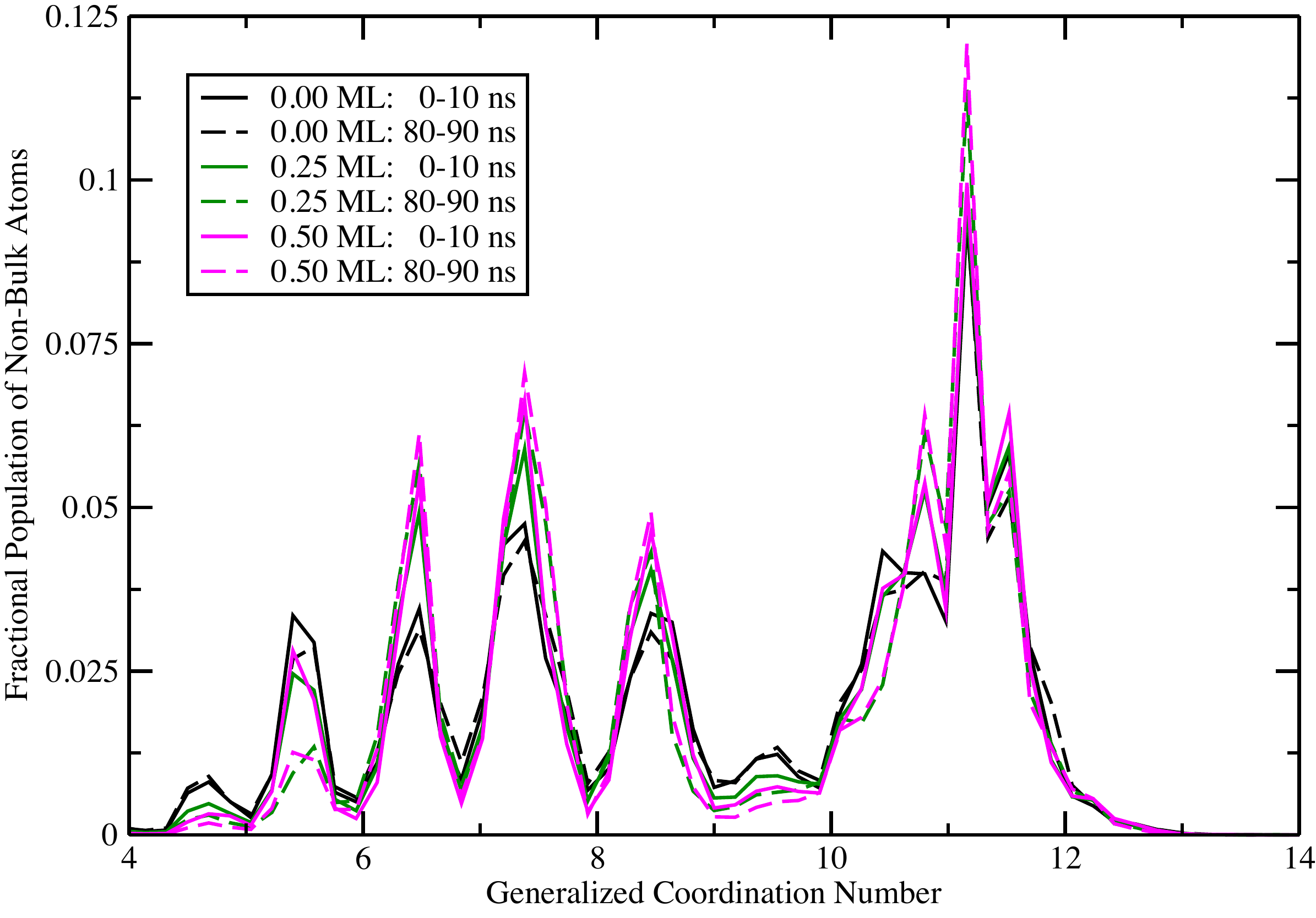}
  \caption{The distribution of GCN values for the near-surface atoms
    of the \ce{Pt}(112) LS systems.  Colors and line styles are the
    the same as Fig. \ref{fig:LS321GCNF}.  There is a notable loss of
    population at GCN $\sim$ 4.5 and 5.5 on the CO-dosed surfaces and
    corresponding increases at 6.5, 7.4, and 8.5.  These changes point
    to a CO-induced flattening of the Pt(112) surface.}
\label{fig:LS112GCNF}
\end{figure}

While the majority of the other surfaces examined in this study
experienced significantly less reconstruction, the Pt(112) LS systems
did exhibit both edge wandering and reconstruction. Figure
\ref{fig:LS112GCNF} shows the evolution of GCNs for the \ce{Pt}(112)
LS systems. There is a similar reduction of population around
GCN$\sim$4.5 and 5.5 which is consistent with both reconstruction and
step \textit{sinking}. The increased peak heights at 6.5, 7.4, and 8.5
indicate a CO-induced flattening of the surface. Representative GCN
plots of the other interfaces are included in the SI as Figures
S2 through S4.

\subsection{Edge wandering}
\subsubsection{Pt (557)}
The Pt(557) surfaces have been explored more fully in previous
experimental and theoretical work,\cite{Tao:2010aa,Michalka:2013aa}
but the step doubling appears to be sensitive to the details of the
Pt-CO interaction. As in previous simulations, there were numerous
instances of edge wandering.  In this work, however, no doubling
nucleation sites formed on the (557) surfaces, although they were seen
previously in Ref. \citenum{Michalka:2013aa}. Our \ce{Pt\bond{-}CO}
interaction has been reduced in binding strength compared with the
previous work, and this may have reduced the driving force for step
doubling.  It is also possible that nucleation of the doubling
reconstruction is a relatively rare event, and 100 ns of simulation
time is insufficient to observe the doubling on Pt(557) at this
temperature.

\subsubsection{Pt (112)}
The Pt(112) MS surfaces exhibited a small amount of CO-induced
restructuring, although this interface exhibits a mode of releasing
surface tension in which some of the \{100\} step edges sink into the
surface, shifting sideways to display a ``sunken'' or lowered \{111\} step
edge. This is highlighted in Figure S5 in the
SI. Adatom formation from the sunken steps was minimal and the total
adatom movement on these systems was reduced compared with interfaces
with wider plateaus.

\begin{figure}
\centering
  \includegraphics[width=0.8\linewidth]{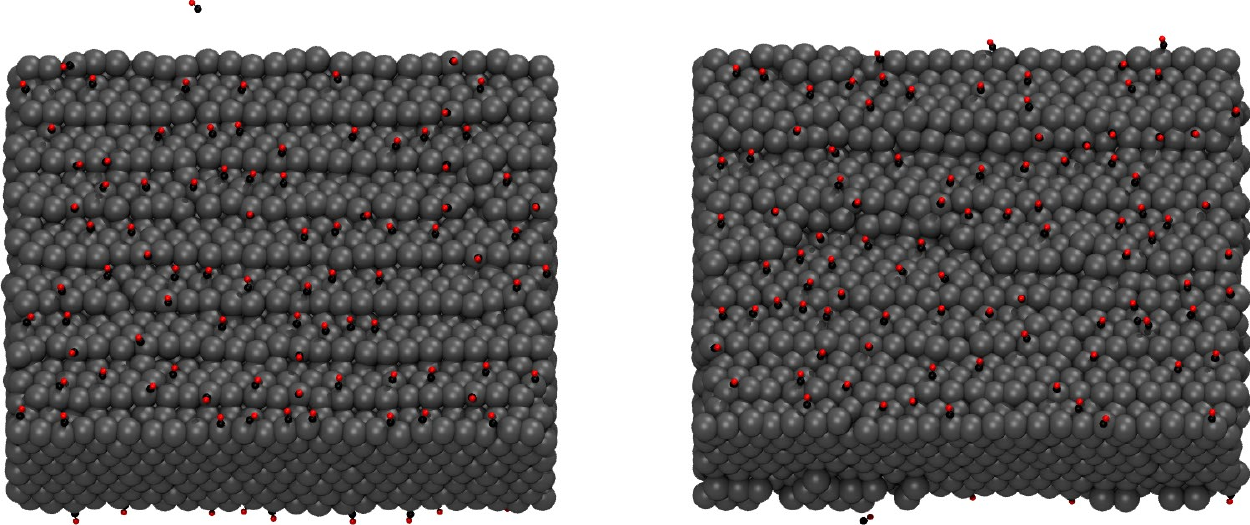}
  \caption{The Pt(112) 0.25 ML LS surface.  At the beginning of the
    simulation (left), minimal step-wandering has occurred. Step
    doubling (near the top of the right panel) and edge sinking (near
    the bottom) both play a role in lowering the surface energy of
    this system.}
  \label{fig:112_25CO}
\end{figure}

The Pt(112) LS surfaces exhibited significant step-wandering, along
with a moderate amount of edge reconstruction. About half of the
steps on each system sank into the bulk, preventing most adatom
formation, but the remaining steps were active sources of adatoms and
edge wandering. Step doubling was observed for the 0.25 ML and
0.5 ML surfaces. Figure \ref{fig:112_25CO} highlights the 0.25 ML LS
system and the step doubling that was observed, along with the
sinking of other step edges on the surface.

\subsubsection{Pt (765)}
The Pt(765) surfaces, while undergoing significant 
step-wandering, did not exhibit step doubling during the
simulations. The MS systems (those extended to display more steps)
maintained the (765) motif over the entire simulation. What appeared
to be the initial stages of a doubling event resulted in portions of
the two edges sinking into the surface of the metal.

The LS systems also experienced significant step wandering. While no
doubling was observed, connections between separate edges were
seen. However, these connections led to step rotation on the surface
of the metal. Despite the apparent stability of the (765) motif for the
MS systems, the 0 and 0.5 ML LS systems underwent a relatively
unstructured surface morphology change. Notably, there is an increase
in the GCN population near 6.8 highlighted in Figure
S4 in the SI.

\subsubsection{Pt (321)}
The small plateau widths and kinked step edges on the Pt(321) surface
showed the most interesting behavior after exposure to CO. Numerous
step-doubling events were observed on both the MS and LS systems with
0.5 ML CO. The 0.25 ML systems also showed significant reconstruction,
although fewer step doubling events were observed.  At the relatively
high temperatures used in this study, the 0 ML systems also
experienced a small amount of step wandering.

\begin{figure}
\centering
\includegraphics[width=0.7\linewidth]{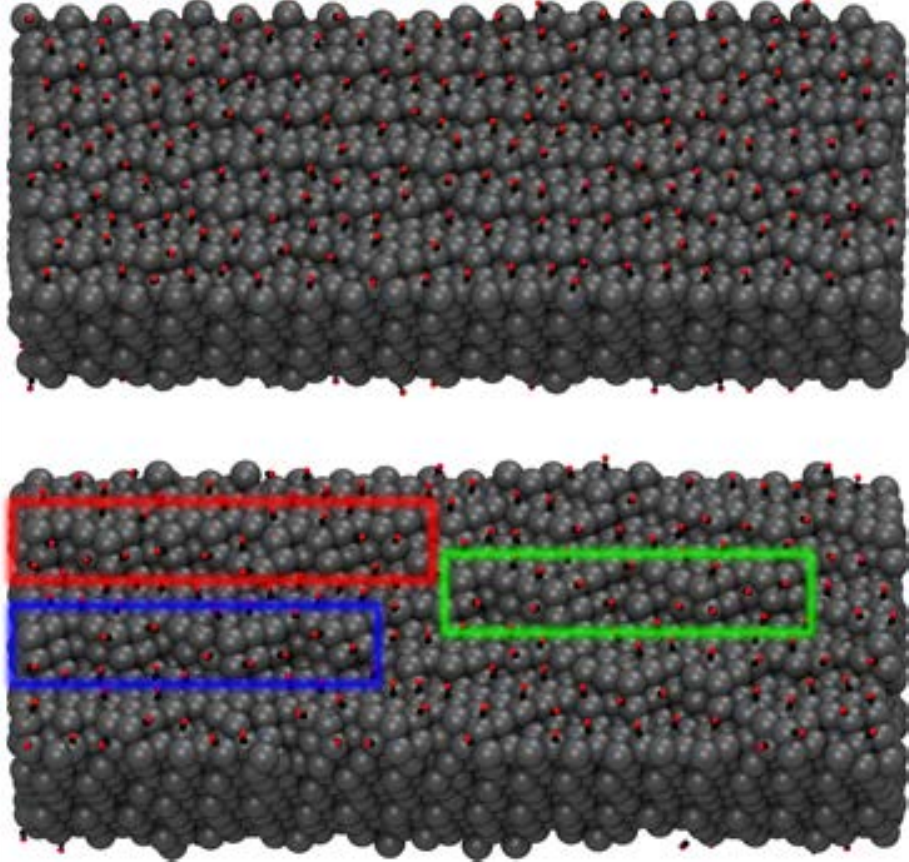}
\caption{The Pt(321) 0.5 ML LS system.  Immediately after dosing with
  \ce{CO} (top), the kinked edges are stable. By 100 ns after dosing,
  the adsorbed \ce{CO} has induced double step formation. The plateaus
  are sufficiently narrow that the formation of double step nucleation
  sites can happen in many locations.  Zippering of the double layers
  completes a reconstruction on other surfaces, but is hindered by the
  presence of a nascent double layer on an adjacent step.  Frustrated
  double layers were common on this surface.}
\label{fig:partialDoubleLayer}
\end{figure}

An interesting case of frustrated double layer formation is
highlighted in Figure \ref{fig:partialDoubleLayer} where three double
layers have nucleated, but the doubling highlighted in green is
preventing both the blue and red double layers from coalescing. Within
the 100 ns encompassed by the simulations, the 0.5 ML LS systems were
not able to form a double step that traversed the length of the
system, while numerous examples of partial or frustrated double steps
were found.

\begin{figure}
\centering
\includegraphics[width=0.7\linewidth]{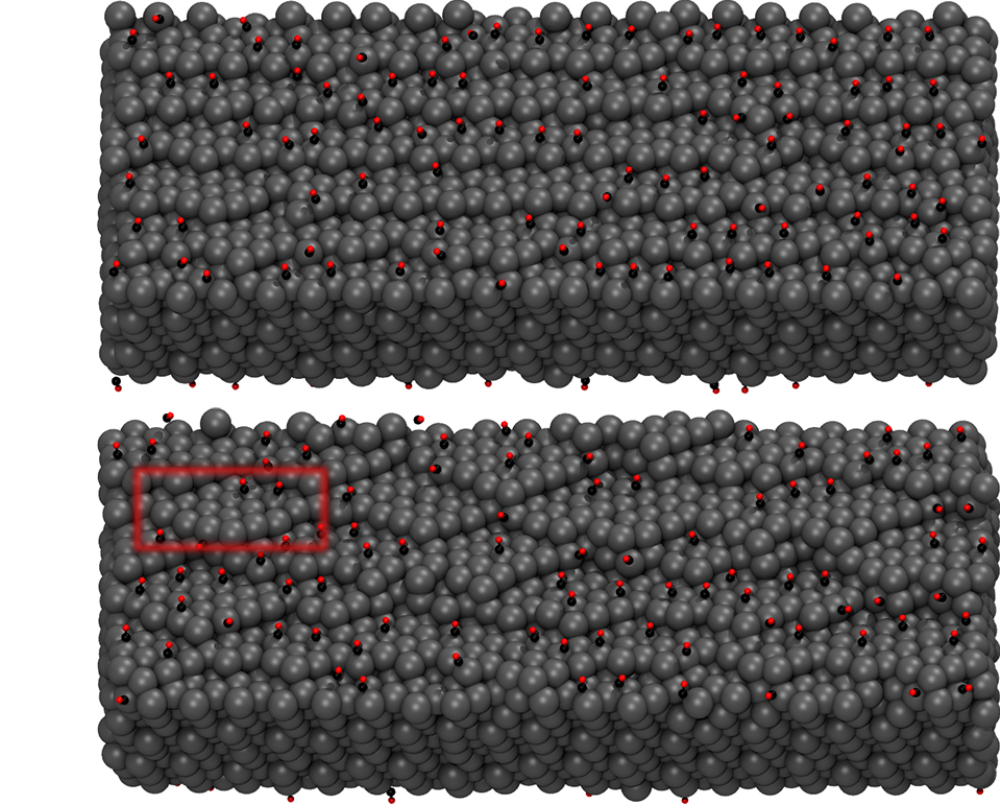}
\caption{The Pt(321) 0.25 ML LS system. Immediately after dosing (top)
  the kinked edges are stable.  After 100ns (bottom), the red box
  highlights one of the repeated (111) ``diamond'' domains formed
  between the initially regular steps.}
\label{fig:diamonds}
\end{figure}

At a lower \ce{CO} coverage (0.25 ML), the Pt(321) surface exhibited
the structure highlighted in Figure \ref{fig:diamonds}. There is a
small degree of step-edge doubling in these systems, but the majority
of this surface exhibited diamond shaped (111) domains.

\section{Discussion}
\subsection{Plateau width}
On these stepped surfaces, the width of the (111) plateaus provides a
good estimate of the surface energy of the facet.  The plateau width
also has an effect on the mechanism of restructuring. The doubling
mechanism proposed in Ref. \citenum{Michalka:2013aa} depends on two
step edges meeting and forming a stable nucleation site.  After
nucleation, a zippering process accelerates the rest of the step
reconstruction.  Because the meeting of two step edges is a stochastic
process, any feature that makes the initial meeting less likely,
e.g. increasing the distance between edges, will make the
reconstruction process more difficult to capture in reasonable
simulation times.

Conversely, a small plateau width should increase the number of
nucleation events, which should in turn lead to increased step
doubling.  Both the (321) and (112) systems show evidence of this.
The Pt(112) LS systems experienced four clear instances of step
doubling -- two on the 0.25 ML LS system, and two on the 0.5 ML LS
system -- all of which were completed within the first 27 ns of the
simulations. The Pt(321) systems, while exhibiting partial double
layers, often appeared in a frustrated configuration, and only three
full ($>$ 95\%) doublings were observed during the simulations.  While
the process of step doubling on the (112) systems occurred fairly
smoothly, the doubling of the (321) surfaces did not. The diamond
motifs highlighted in Figure \ref{fig:diamonds} were stable for the
majority of the simulation.  In the instances where the diamonds broke
apart, this was followed by an adjacent step doubling within $\sim$5
ns. Thus, despite nucleation sites forming rapidly on the (321)
surfaces, the stability of these sites slowed or even prevented the
step doubling process from going to completion.


\subsection{Energy to separate from an edge}
\begin{figure}[p!]
  \centering
  \includegraphics[width=0.9\linewidth]{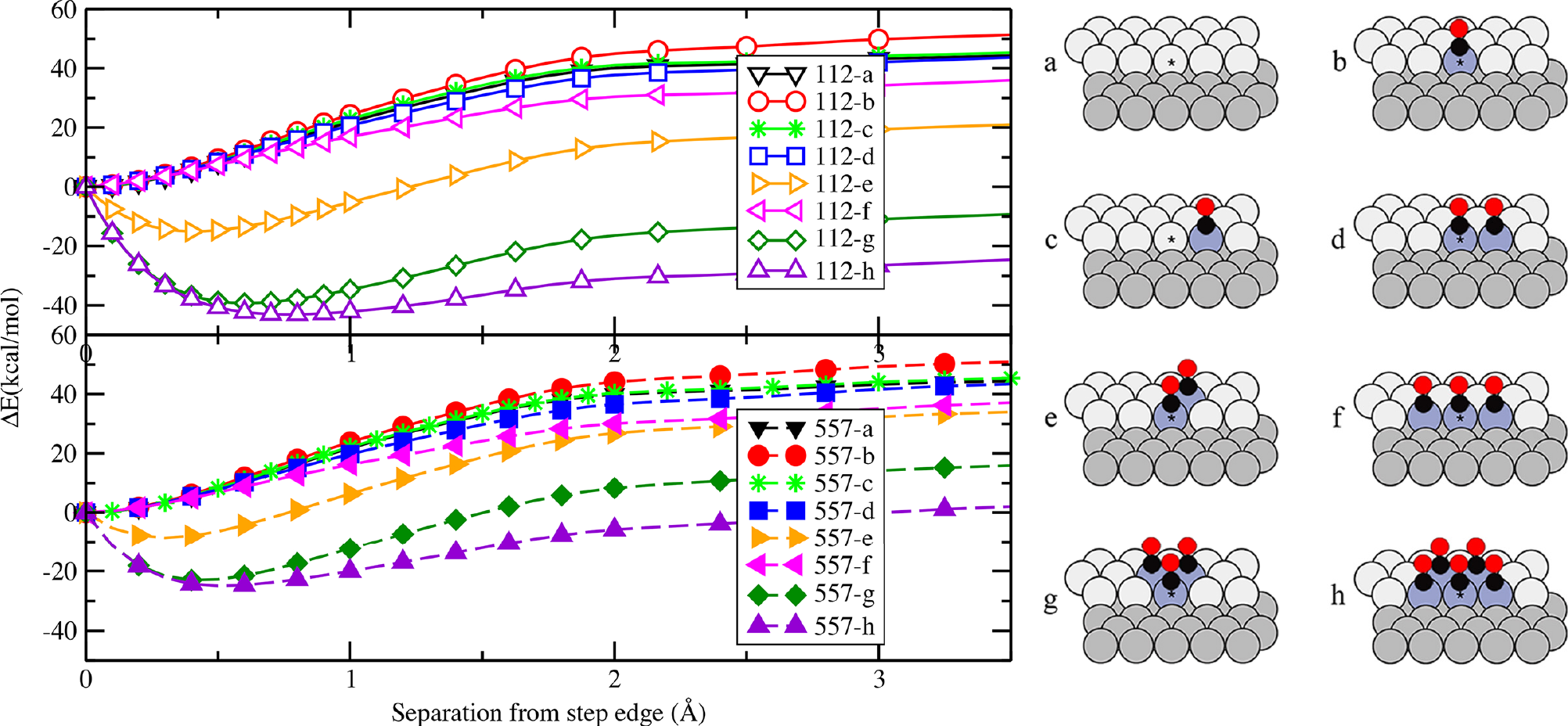}
  \caption{Energies for displacing an edge atom (*) perpendicularly
    from the (112) (top) and (557) (bottom) step edges.  Each of the
    energy curves corresponds to one of the labeled configurations on
    the right and are referenced to the unperturbed step edge. The
    spheres represent Pt atoms on the upper (white) and lower (grey)
    steps.  Colored atoms (blue) are depicted with a CO molecule
    adsorbed in an atop configuration. Certain configurations of CO,
    notably $e$, $g$ and $h$, can lower the energetic barrier for
    creating an adatom.}
\label{fig:112_557_ES}
\end{figure}

\begin{figure}[p!]
  \centering
  \includegraphics[width=0.9\linewidth]{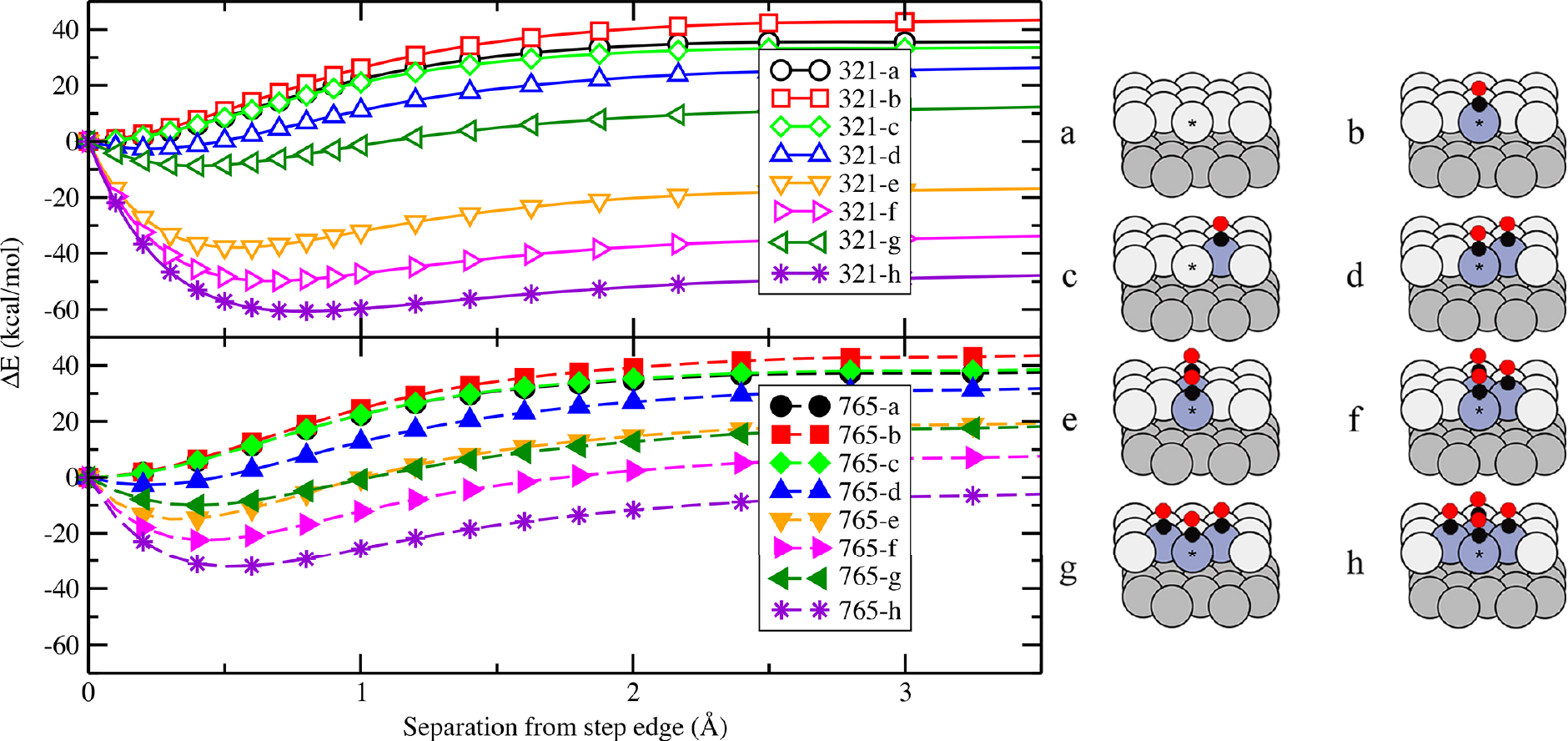}
  \caption{Energies for displacing an edge atom (*) perpendicularly
    from the (321) (top) and (765) (bottom) kinked step edges.  The
    configurations to the right of the graphs show the kinked step
    edges of these facets. The kinked steps lower the energetic
    barrier for adatom creation when compared to the flat steps of the
    (112) and (557) surfaces. Plateau width also has a significant
    effect on the energetic benefit of adatom formation.}
\label{fig:321_765_ES}
\end{figure}

Adatom creation is essential for step wandering and doubling.  To
show the effects of \ce{CO} adsorbates on this process, we have
computed potential energies for adatom creation under several \ce{CO}
configurations.\cite{Michalka:2013aa,Michalka:2015aa} These energy
surfaces were computed by displacing one Pt edge atom along the
plateau in a direction perpendicular to the step edge (shown in Figs.
\ref{fig:112_557_ES} and \ref{fig:321_765_ES}). The strong quadrupolar
repulsion between bound \ce{CO} molecules makes adatom formation
energetically favorable at high coverages. For the (112) and (557) facets,
configurations $e$, $g$, and $h$, shown in Fig. \ref{fig:112_557_ES},
make the initial formation of an adatom an energetically favorable
process. Because the step edges are kinked in the (321) and (765) facets,
more \ce{CO} configurations ($e$, $f$, $g$, and $h$) are favorable for
adatom formation (see Fig. \ref{fig:321_765_ES}).

This analysis provides several useful bits of data. First,
removing an atom from a (321) or (765) step is more energetically
favorable than from a (112) or (557) step.  The former two facets have
kinked step edges, which lower the coordination number for the edge
atoms. This trend is clearly shown in Figures \ref{fig:112_557_ES} and
\ref{fig:321_765_ES} for configuration $a$ where no CO is present
on the surface. In these instances, creating an adatom from a kinked
step edge is 10 kcal/mol more favorable than from a flat step.

Additionally, the width of the plateaus affects the energy required to
form an adatom at higher coverages.  Configurations $e$, $f$, and $h$
in Figure \ref{fig:321_765_ES} for the kinked steps highlights this
result. It is approximately 25 kcal/mol more favorable to form an
adatom on the (321) surface than it is on the (765)
surface. Similarly, there is approximately a 20 kcal/mol difference in
favor of forming an adatom from a (112) step-edge when compared to the
(557) surface.

Finally, the location and arrangement of \ce{CO} strongly affects the
energy required to remove the atom from the step edge. Separation
becomes most favorable when the candidate adatom and the atom directly
behind it both have an atop-adsorbed \ce{CO}. This configuration
directs the strong quadrupolar repulsion directly away from the step
edge, lowering the barrier for adatom formation.  Comparing
configurations $d$ and $e$ in both figures illustrates this effect. In
configuration $e$, both adsorbed \ce{CO} molecules are situated along
a line which is perpendicular to the step edge.  In configuration $d$
the repulsion due to the \ce{CO} is mostly parallel to the step
edge. As the coverage increases, the likelihood increases for
observing one of the \ce{CO} configurations conducive to adatom
formation, thereby increasing Pt surface mobility.

\subsubsection{Edge Ordering}
The steps in the (321) facet were observed to change from kinked edges
to larger \{100\} edge segments.  After dosing with \ce{CO}, the change in
angular ordering of the steps was analyzed to track the straightness
of step edges for all four facets. Assuming atoms $i-1$, $i$, and
$i+1$ are sequential along a step edge, $\mathbf{a}_{i}$ is defined as
the vector between $i-1$ and $i$ and $\mathbf{b}_{i}$ is the vector
between $i$ and $i+1$. An effective measure of the ordering is,
\begin{equation}
\left< \cos^2 \theta \right> =  \left< \frac{1}{N_\mathrm{edge}} \sum_i
  \left(\frac{\mathbf{a}_{i}\cdot\mathbf{b}_{i}}{\left|\mathbf{a}_{i}\right|\left|\mathbf{b}_{i}\right|}\right)^2 \right>,
\label{eq:order}
\end{equation}
where the averaging is done over all of the edges present in a
configuration. Here $N_\mathrm{edge}$ is the number of atoms present
in the edge.  Atoms in straight step edges would result in
$\cos^2\pi = 1$ while the ideal (321) kinked surface would result in
$\cos^2 \left(\frac{2 \pi}{3}\right) = 0.25$.

\begin{figure}
  \includegraphics[width=\linewidth]{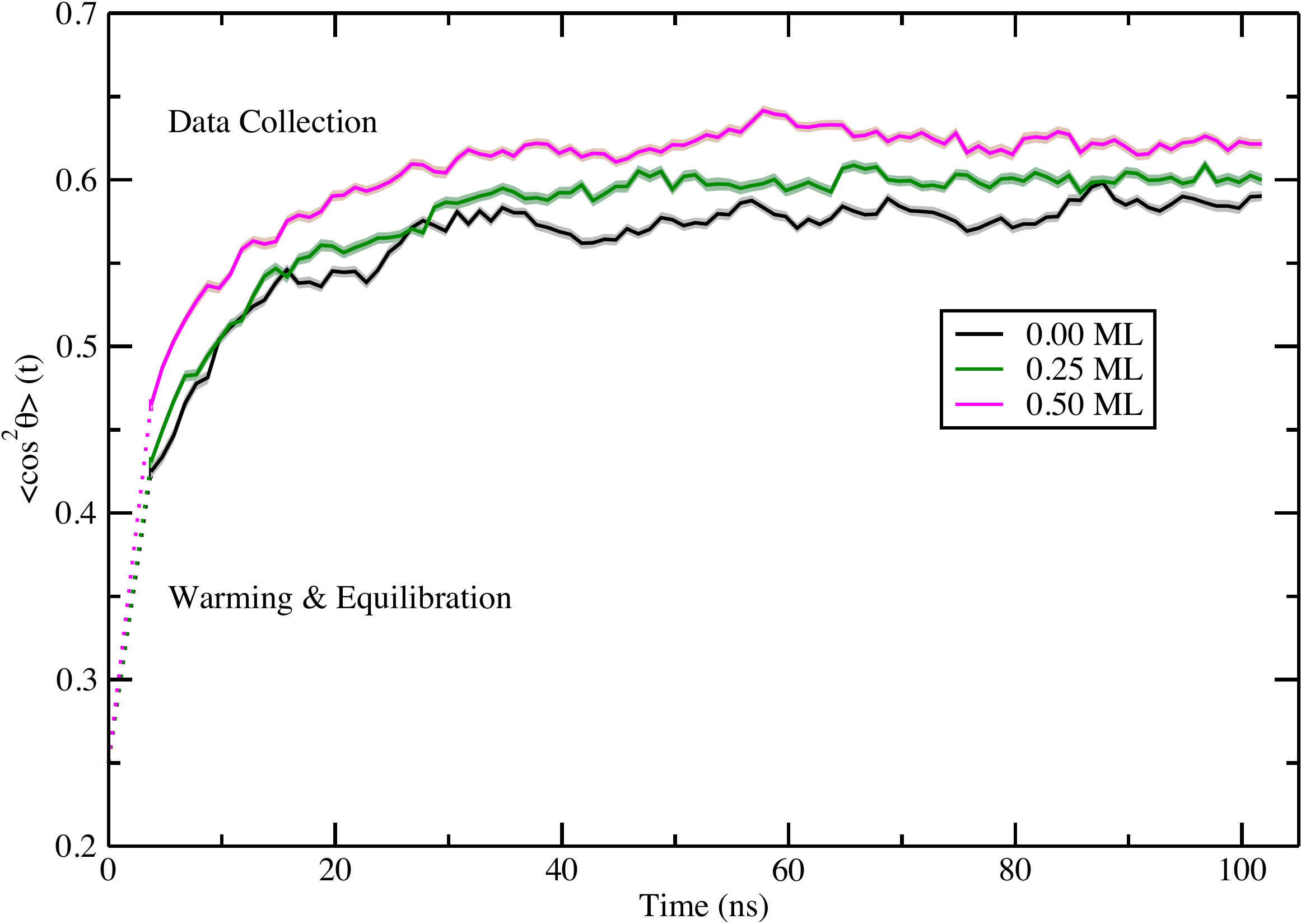}
  \caption{Evolution of the edge-ordering over time for the Pt(321)
    systems.  Both the bare and CO-dosed surfaces start near the ideal
    $\left<\cos^2\theta\right> = 0.25$ and rapidly rise during the
    first 40 ns of the simulation. Generally, the presence of \ce{CO}
    leads to increased linear ordering of the step-edges.}
  \label{fig:eo}
\end{figure}

For the (321) facet, the systems start near the ideal
$\cos^2\theta = 0.25$ but quickly grow towards a more ordered set of
edges (see Fig. \ref{fig:eo}). Some ordering by \ce{CO} coverage is
observed, where the systems with \ce{CO} tend to display more ordered
edges.  This appears to be due both to the lower energy of the \{100\}
edge segments and the CO-induced mobility of edge atoms.



\subsection{Mechanisms of structural changes}
Our previous work on \ce{Pt}(557) revealed that adsorbed \ce{CO}
destabilizes the step-edges leading to rapid formation of adatoms and
step-wandering on the surface. If sufficient step-wandering occurs,
two step edges can connect to form a nucleation site. Once one of
these sites is formed, a zippering of the double step across the
length of the simulation cell happens rapidly.

This mechanism depends on the stochastic meeting of two step edges
which is correlated with the density of metal adatoms on the surface.
Thermal energy can increase the number of adatoms, however in the
current work, this pathway is limited due to lower simulation
temperatures. The systems without CO, while exhibiting some adatom
formation and re-absorption, showed minimal reconstruction. When
\ce{CO} was introduced, moderate surface reconstruction was
observed. The presence of \ce{CO} plays a similar role to increased
temperature by weakening the \ce{Pt\bond{-}Pt} bonds along the
step-edge.

Besides the density of adatoms, another factor in step edge wandering
is the plateau width. The narrow plateaus in the (112) and (321)
systems makes step edge meetings more likely.  In previous work on the
(557) surface, only one nucleation site was necessary for a step
doubling. For many of the surfaces explored in this study, many
nucleation sites are created before the steps complete the
doubling. In some cases, this leads to frustrated doubling,
highlighted in Fig. \ref{fig:partialDoubleLayer}.

All of the (321) surfaces, even those without \ce{CO} present, showed
some amount of surface roughening and reconstruction suggesting that
these surfaces have limited stability at high temperatures. However,
the presence of 0.25 and 0.5 ML coverages of \ce{CO} did lead to
significantly more step-wandering, intermediate reconstructions, and
partial and full double layer formation.

\section{Summary}
The mechanism and dynamics of reconstruction on \ce{Pt} surfaces,
specifically step doubling, depends on the displayed surface facet. It
is also dependent on the presence of adsorbates, here \ce{CO}, that
can disrupt the step-edges and then stabilize the formation of double
steps. The energetics of step-edge breakup shows that the kinked
surfaces, (321) and (765), are especially favorable for formation of
adatoms due to undercoordination of the edge atoms.

The mechanism proposed in our earlier work involving stochastic
edge-doubling nucleation followed by zippering is mostly supported by
the current work.  Specifically, systems with narrower plateaus form
frustrated double layers because multiple nucleation sites form before
zippering can be completed.

Re-parameterization of the \ce{Pt\bond{-}CO} interaction to match new
experimental and DFT binding energies has altered the dynamics of the
step-doubling process, particularly for the (557) surfaces. While
significant step-wandering was observed on Pt(557), the density of
adatoms was not sufficient to form nucleation sites during the 100 ns
simulation time.

One notable new finding is the observation that on the (321) facet, the
CO adsorbate can assist in restructuring the kinked step edge into
longer segments of straight \{100\} edges.  With high CO coverages, the
kink atoms are highly susceptible to ejection from the edge, but they
can add to a straight segment with a lower energetic penalty.  This
energetic argument indicates that kinked steps, like the ones
exhibited by the Pt(321) and Pt(765) surfaces, should be most
susceptible to CO-induced reconstruction.

The cell formed by step-edge atoms may also play a role in the
dynamics of reconstruction. Tao {\em et al.}\cite{Tao:2010aa},
examining both the Pt(557) surface that exhibits a (100) step edge and
the Pt(332) surface that exhibits a (111) edge, observed that the
extent and final structure of the \ce{CO}-induced reconstruction were
affected by the type of step edge. The Pt(557) surface showed a
significant degree of reconstruction at low pressures, and was
observed to form triangular nanoclusters at high coverages. The
Pt(332) surface exhibited less reconstruction at low CO coverages and
formed rectangular nanoclusters at high coverages. Investigation of
the detailed molecular dynamics of (100)-edge surfaces like Pt(332)
will increase our ability to predict the dynamics of reconstruction on
these surfaces, and will be the topic of future study.

Recent work on \ce{CO}-induced nanostructure formation on a
\ce{Cu}(111) surface has shown that when the metal-metal binding
interactions are weaker, the presence of strong repulsive adsorbates
can be sufficient to disrupt even flat surfaces.\cite{Eren:2016qt} We
have reached a similar conclusion with the \ce{Pt} systems in our
simulations.  Surfaces that have a higher surface energy and more
sites where adatoms can be easily formed are more likely to experience
reconstruction in general.  In particular, step doubling increases the
fraction of the surface that is occupied by the low surface-energy
(111) domains.

\section{Supporting Information}
The Supporting Information includes: GCN data for an additional ideal
surface (Pt(557)) to complement the Pt(321) surface in
Fig. \ref{fig:ideal321GCN}, distributions of GCN values for the
Pt(557) LS and Pt(765) systems, Pt(112) step edge configurations, and
sunken edge illustrations for the Pt(765) system.

\begin{acknowledgement}
  Support for this project was provided by the National Science
  Foundation under grant CHE-1362211 and by the Center for Sustainable
  Energy at Notre Dame (cSEND). Computational time was provided by the
  Center for Research Computing (CRC) at the University of Notre Dame.
\end{acknowledgement}

\newpage
\bibstyle{achemso}

\providecommand{\latin}[1]{#1}
\providecommand*\mcitethebibliography{\thebibliography}
\csname @ifundefined\endcsname{endmcitethebibliography}
  {\let\endmcitethebibliography\endthebibliography}{}

\begin{tocentry}
\center\includegraphics[height=1.75in]{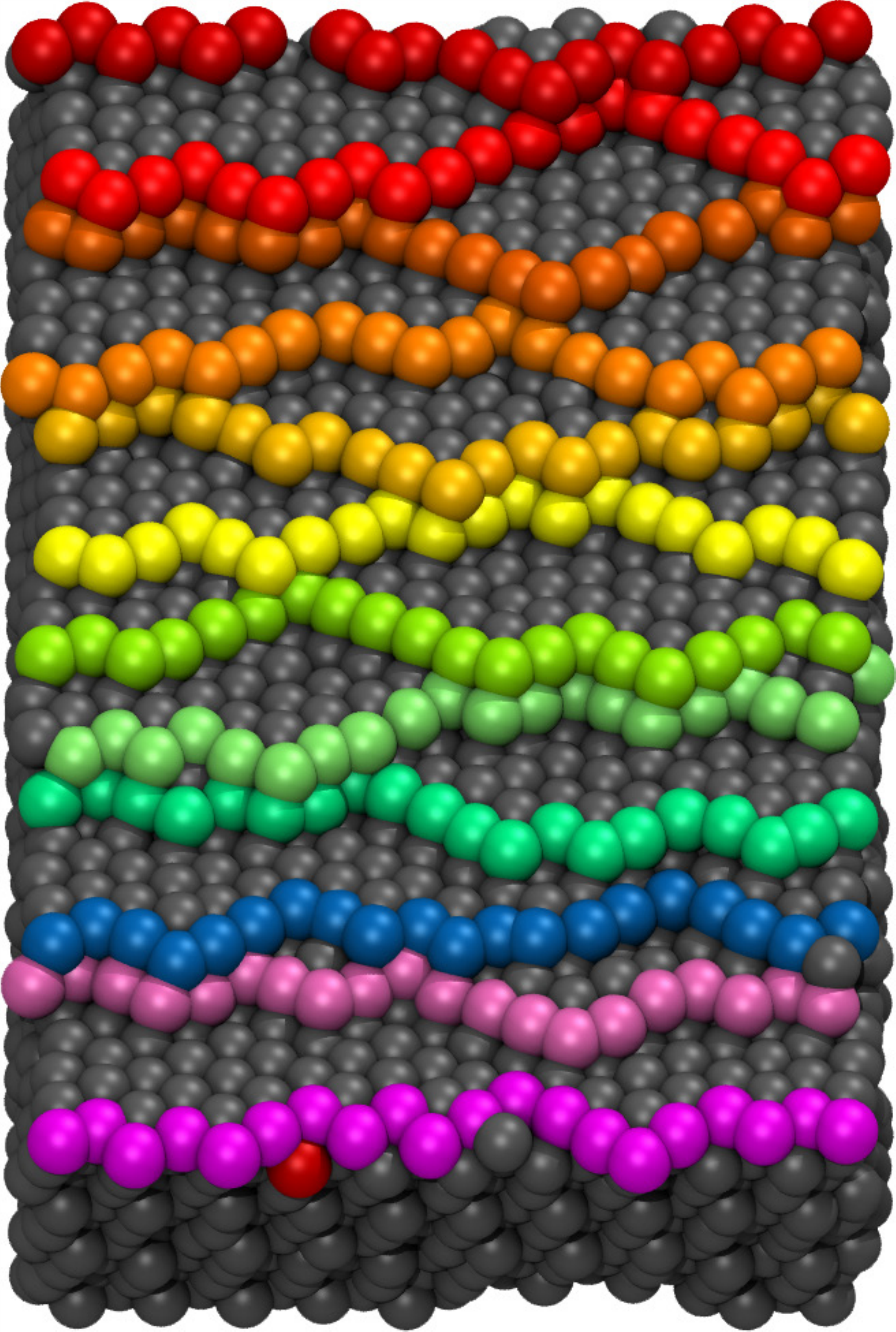} 
\end{tocentry}

\end{document}

\begin{figure}
\centering
\includegraphics[width=0.9\linewidth]{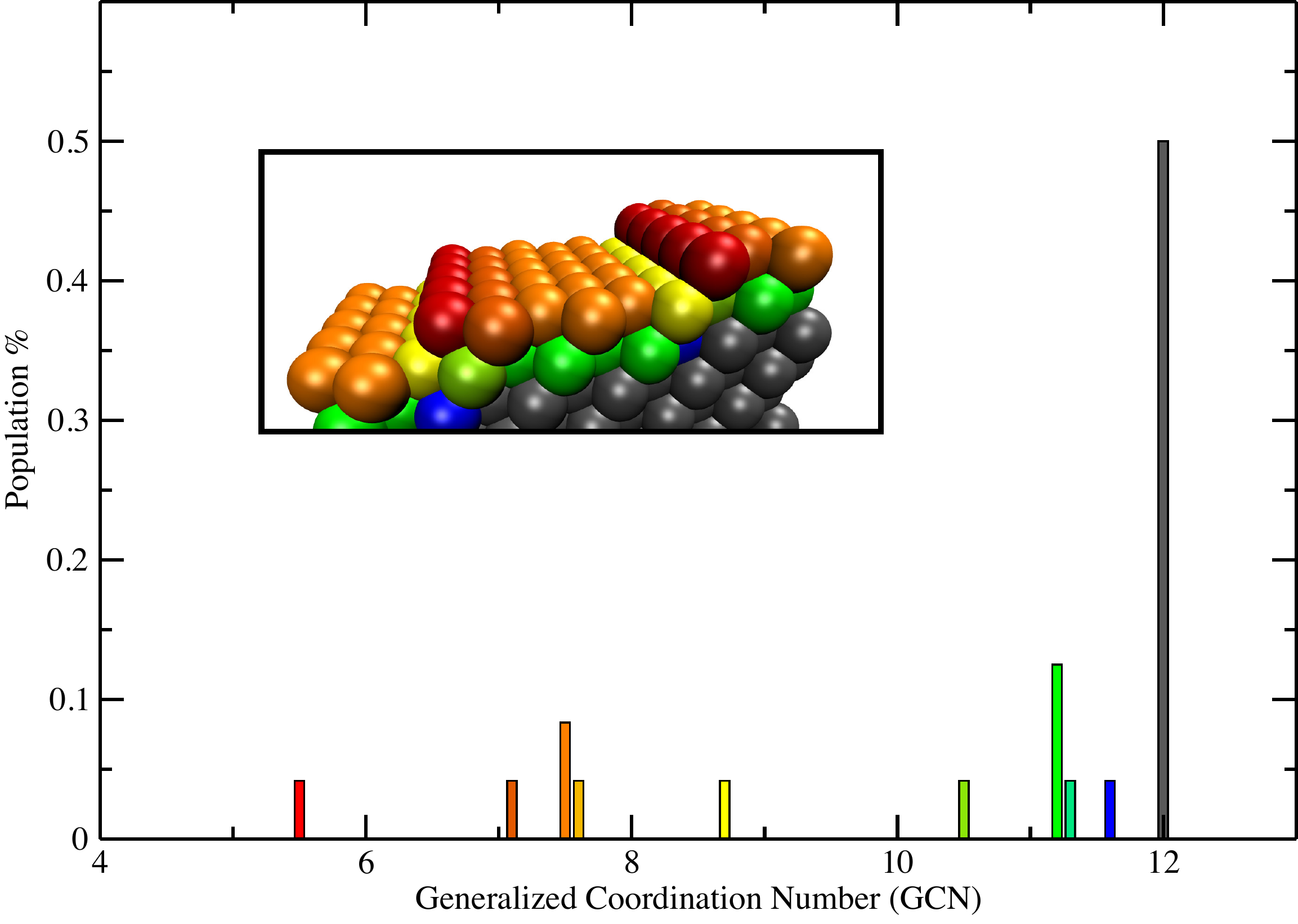}
\caption{The generalized coordination number of an ideal Pt(557)
  system colored to match the inset structure. Other than the bulk
  (GCN = 12), the ideal (557) surface displays a variety of coordination
  environments at its surface. The under-coordinated edge atoms (red)
  have the lowest GCN, where the plateau atoms have GCN values around
  the ideal (111) surface number of 7.5.  Subsurface atoms (yellow,
  greens, and blues) have a larger GCN than the surface, but are still
  less coordinated than the bulk.}
\label{fig:557GCN}
\end{figure}

Step-wandering during the simulations can interfere with easy
interpretation of the generalized coordination figures. Figure
\ref{fig:557GCN} provides an additional ideal surface to complement
the ideal Pt(321) GCN distribution shown in the main paper. The Pt(557)
surface has straight step edges, and the GCN distribution therefore
has fewer peaks than the Pt(321) surface.

\newpage

\begin{figure}
\centering
\includegraphics[width=0.9\linewidth]{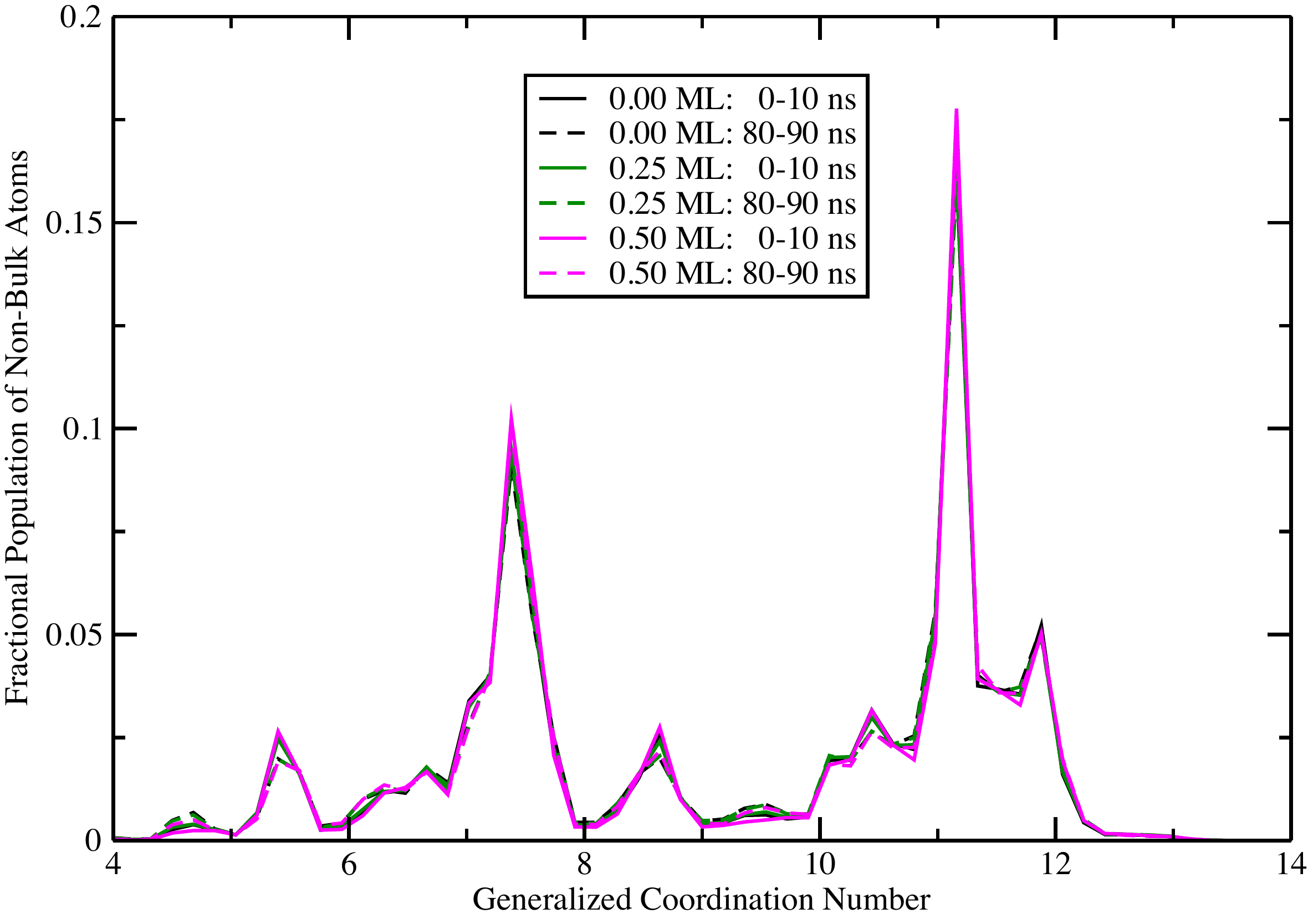}
\caption{The distribution of generalized coordination numbers (GCN)
  for the near-surface atoms of the \ce{Pt}(557) LS systems.  Solid
  lines represent the averaged GCN during the first 10 ns of
  simulation time, while dotted lines correspond to the last 10
  ns. Black, green and magenta lines depict the 0, 0.25, and 0.5 ML
  systems respectively. Except for a small decrease at GCN values of
  5.5 and 10.5, minimal changes were observed.}
\label{fig:557lsGCN}
\end{figure}

The \ce{Pt}(557) surface did not exhibit step doubling in the
simulations reported in this work, so the distribution of generalized
coordination numbers did not undergo significant changes, as shown in
Figure \ref{fig:557lsGCN}.

\newpage

\begin{figure}
\centering
\includegraphics[width=0.9\linewidth]{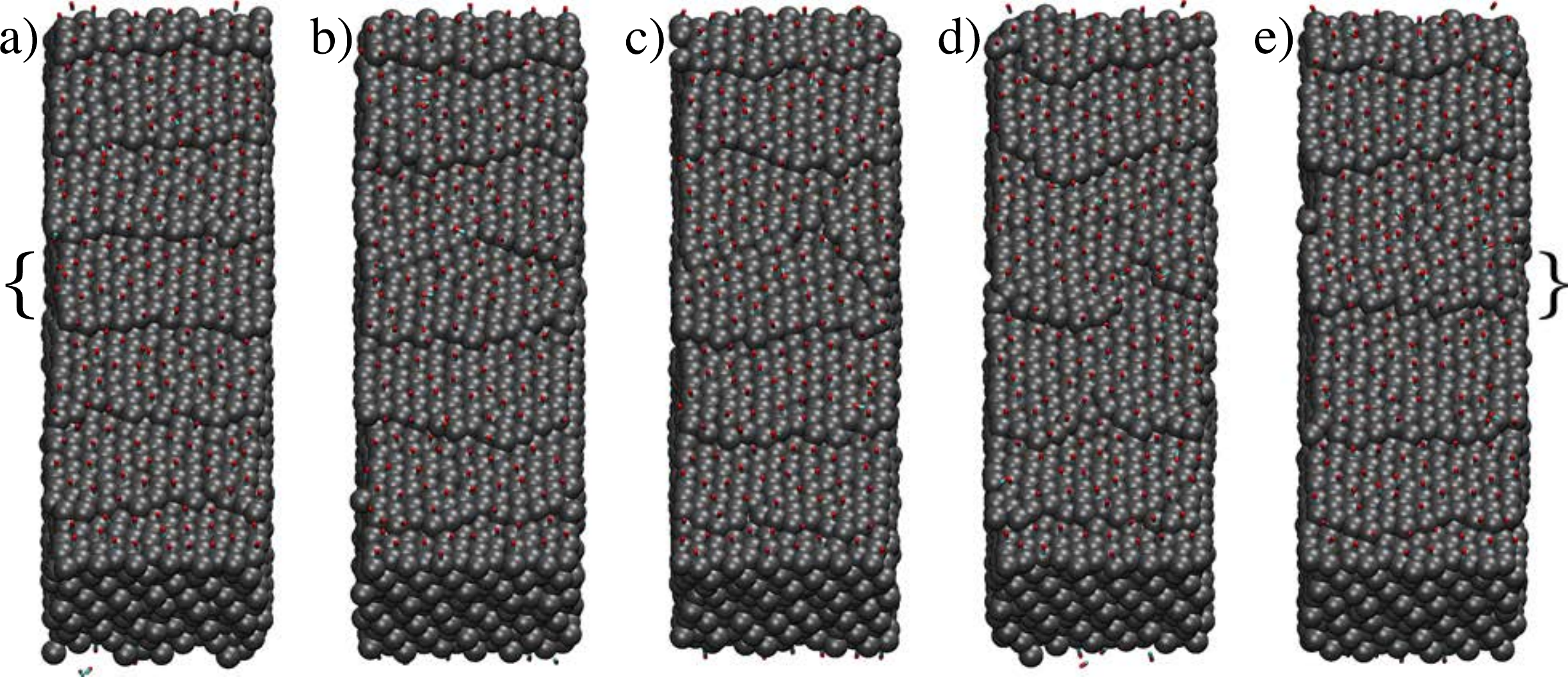}
\caption{The Pt(765)-MS 0.5 ML system (a) 0 ns, (b) 33.4 ns, (c) 50.2
  ns, (d) 75.1 ns, (e) and 100 ns after exposure to CO.  The
  step-edges indicated with a bracket in (a) approach each other while
  sinking into the surface. The result is a step-edge that is between
  the starting points of both parents, but which is only one atom
  elevated from the surface.}
\label{fig:765Edge}
\end{figure}

On the Pt(765) surfaces, reconstruction on only the MS 0.5 ML system
was observed, but none of the other surfaces exhibited reconstruction
beyond step wandering. The slight increase in the peak at GCN
$\sim 7.5$ suggests that this measurement is sensitive to relatively
minor surface reconstruction, like the step-edge doubling and
disappearance highlighted in Figure \ref{fig:765Edge}.

\begin{figure}
\centering
\includegraphics[width=0.9\linewidth]{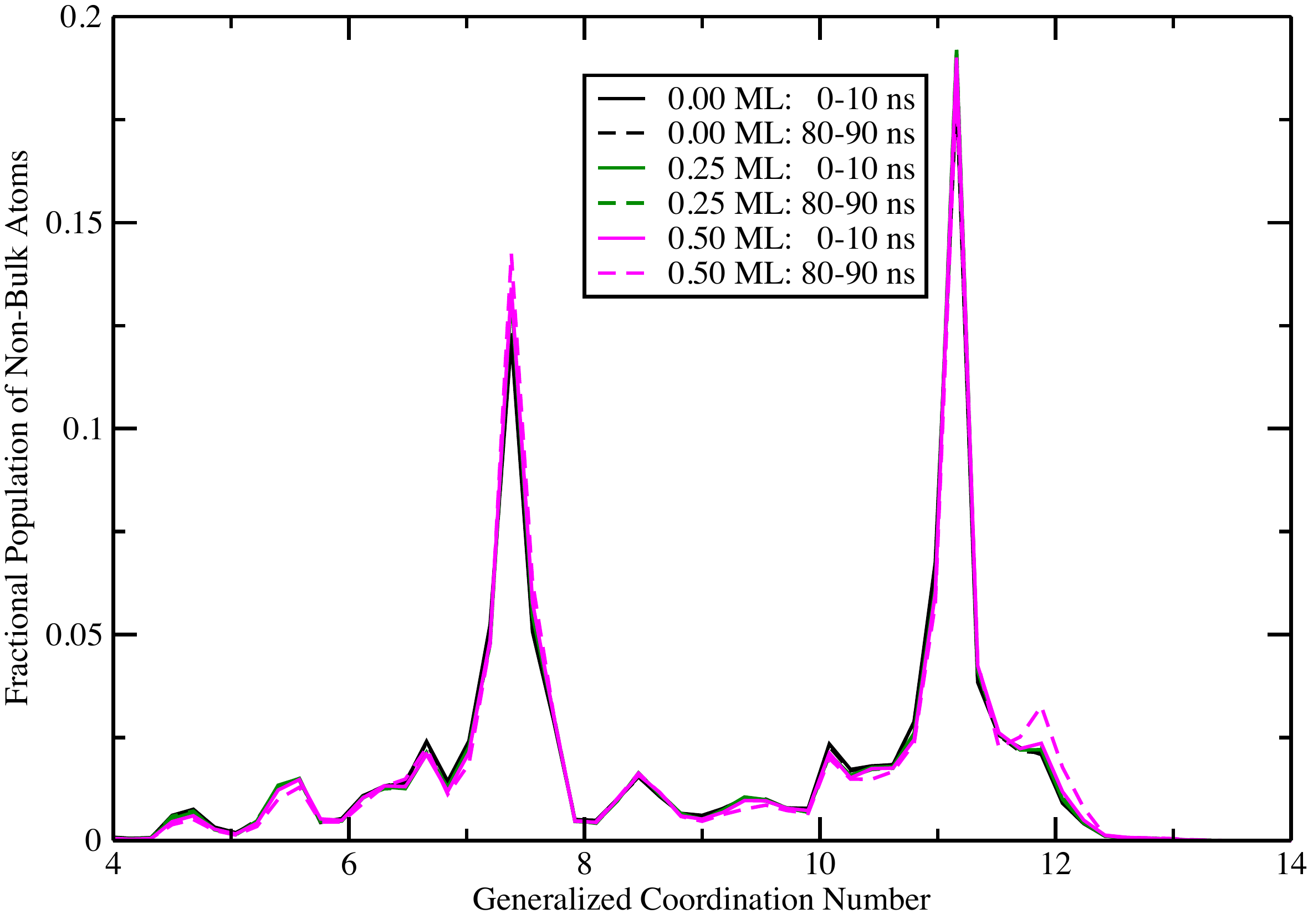}
\caption{The distribution of generalized coordination numbers (GCN)
  for the near-surface atoms of the \ce{Pt}(765) MS systems.  Line
  styles and colors are the same as in Fig. \ref{fig:557lsGCN}. The
  increasing peak heights seen at GCN $\sim$ 7.5 and 11.8 capture the
  doubling process explored in Figure \ref{fig:765Edge}.}
\label{fig:765msGCN}
\end{figure}
\newpage

The increase in the peak height at GCN $\sim$ 6.8 for the 0 and 0.5 ML
LS systems appears to be capturing the loss of the clean (765)
surface. In contrast, the 0.25 ML LS system only exhibited step-edge
wandering, but generally maintained the originally-displayed edges.

\begin{figure}
\centering
\includegraphics[width=0.9\linewidth]{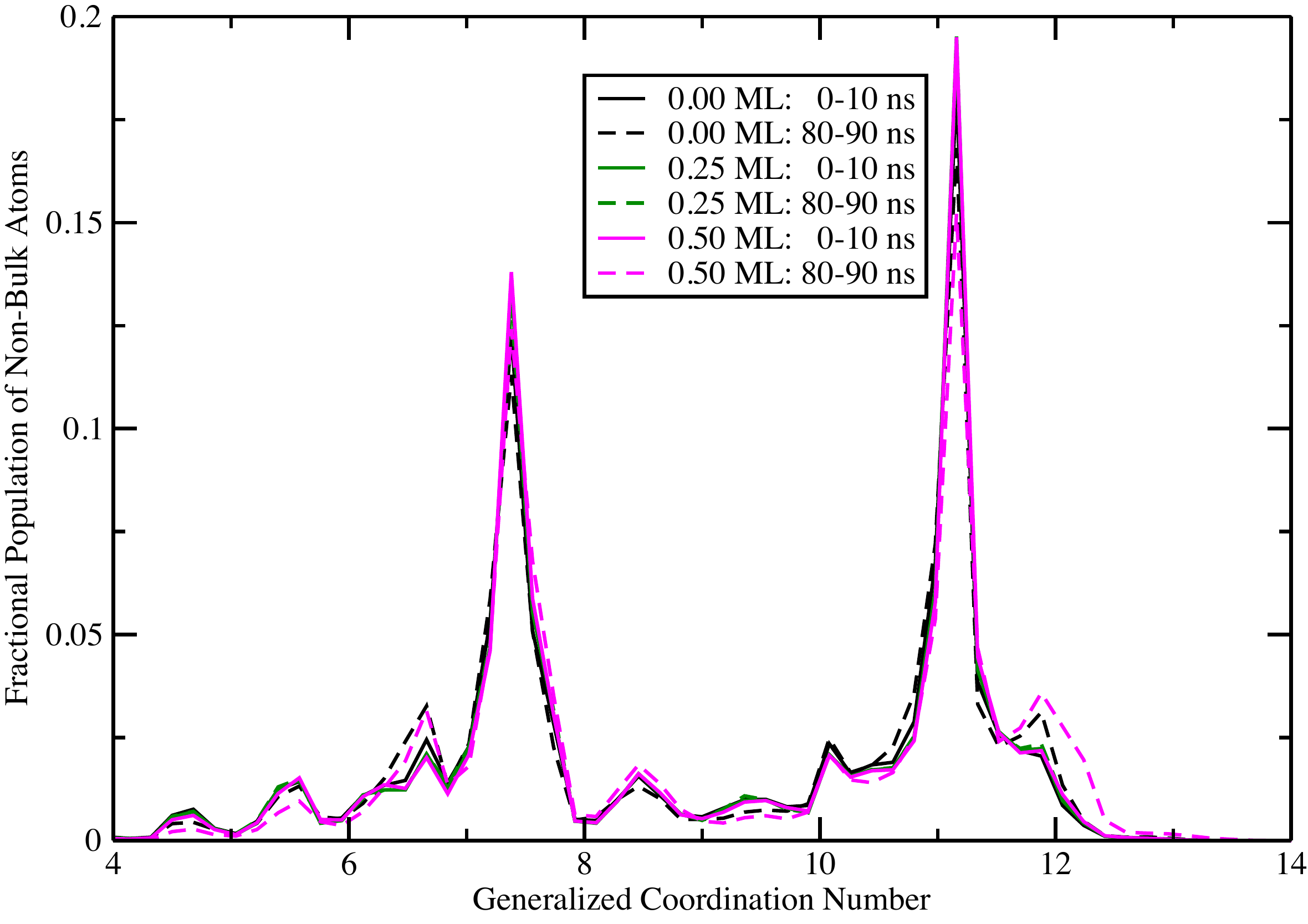}
\caption{The distribution of generalized coordination numbers (GCN)
    for the near-surface atoms of the \ce{Pt}(765) LS systems.  Line
    styles and colors are the same as in Fig. \ref{fig:557lsGCN}. The
    increasing peak heights seen at GCN $\sim$ 6.8 and 11.8 for the
    0.0 and 0.5 ML systems capture the disruption mentioned in the
    main text.}
\label{fig:765lsGCN}
\end{figure}
\newpage

\begin{figure}
\centering
\includegraphics[width=0.3\linewidth]{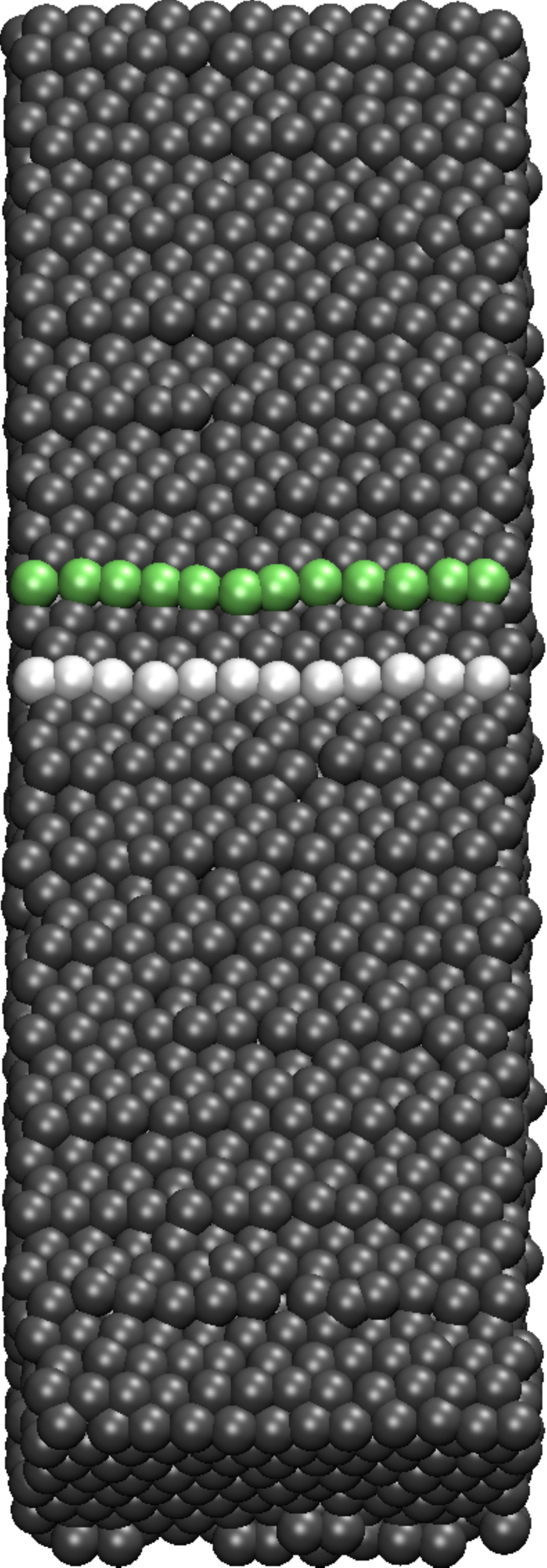}
\caption{The Pt(112) systems have a large surface energy, and
  exhibited restructuring that resulted in many of the \{100\} steps
  sinking into the surface (white) and shifting half a unit-cell to
  display a sunken \{111\} step edge. A few step edges (green) retained
  \{100\} step edge morphology and were the primary source for adatom
  formation and edge wandering.}
\label{fig:112sunken}
\end{figure}

The Pt(112) systems explored in this study were run at temperatures
close to a surface melting transition.  One reflection of this is the
partial sinking of step edges into the surface. In Figure
\ref{fig:112sunken}, a few of the edges maintained \{100\} step facets and
were a source for adatoms. However, the sunken steps rarely ejected
adatoms, and for the majority of the simulation, exhibited minimal
movement on the surface.

\newpage


The kinked edges and large plateaus of the Pt(765) systems help
distinguish which surface attributes were most important in
encouraging or hindering surface reconstruction. No clear
step-doubling occurred on these surfaces despite a significant amount
of step-wandering. On the 0.5 ML MS system shown in Fig.
\ref{fig:765Edge}, what were originally two separate steps, coalesced
into a single step located at an intermediate distance between the
original step edges. A portion of each step sunk into the surface
allowing the two remaining plateaus to meet up to form a single step
edge.

\newpage